\documentclass[%
reprint, twocolumn
showpacs,
preprintnumbers,
 amsmath,amssymb, aps,
floatfix,
pre]{revtex4-1}

\usepackage{graphicx}% Include figure files
\usepackage{dcolumn}% Align table columns on decimal point
\usepackage{bm}% bold math
\usepackage[usenames,dvipsnames]{color} %\usepackage[dvips]{color} 
\usepackage{hyperref}

\newcommand{\changed}[1]{#1}  %Steffen 

\newcommand{\slfrac}[2]{\left.#1\middle/#2\right.}
\renewcommand{\vec}[1]{{\bf{#1}}}
\newcommand{\upd}{\, \mathrm{d}}

\begin{document}

%\preprint{APS/123-QED}

\title{Global attractors and extinction dynamics of cyclically competing species}

\author{Steffen Rulands}
\author{Alejandro Zielinski}
\author{Erwin Frey}
\affiliation{%
Arnold Sommerfeld Center for Theoretical Physics and Center for NanoScience, Physics Department, Ludwig-Maximilians-Universit\"at M\"unchen, Theresienstra{\ss}e 33, D-80333 M\"unchen, Germany
}

\date{\today}

\begin{abstract}
Transitions to absorbing states are of fundamental importance in non-equilibrium physics as well as ecology. In ecology,  absorbing states correspond to the extinction of species. We here study the spatial population dynamics of three cyclically interacting species. The interaction scheme comprises both direct competition between species as in the cyclic Lotka-Volterra model, and separated selection and reproduction processes as in the May-Leonard model. We show that the dynamic processes leading to the transient maintenance of biodiversity are closely linked to attractors of the nonlinear dynamics for the overall species' concentrations. The characteristics of these global attractors change qualitatively at certain threshold values of the mobility, and depend on the relative strength of the different types of competition between species. They give information about the scaling of extinction times with the system size and thereby the stability of biodiversity. We define an effective free energy as the negative logarithm of the probability to find the system in a specific global state before reaching one of the absorbing states. The global attractors then correspond to minima of this effective energy landscape and determine the most probable values for the species' global concentrations. As in equilibrium thermodynamics, qualitative changes in the effective free energy landscape indicate and characterize the underlying non-equilibrium phase transitions. We provide the complete phase diagrams for the population dynamics, and give a comprehensive analysis of the spatio-temporal dynamics and routes to extinction in the respective phases.
\end{abstract}


\maketitle

\changed{Absorbing states play an important role in ecology, where they correspond to the extinction of species~\cite{Frey2010}.} While any stochastic system will eventually end up in one of its absorbing states, in nature, one finds a surprisingly long-term maintenance of biodiversity in ecosystems containing a broad variety of coexisting species. A structured environment in combination with cyclic competition between species was proposed to be a main facilitator of biodiversity~\cite{Durrett1997,Durrett:1998p203}. Classical ecological examples for cyclic interactions comprise coral reef invertebrates \cite{Jackson-1975}, rodents in the high arctic tundra in Greenland \cite{Gilg-2002}, and cyclic competition between different mating strategies of lizards \cite{sinervo-1996-340}. However, long reproduction times and large spatial scales involved make it difficult to quantitatively analyze these ecological systems. To circumvent these problems, recent experimental studies have turned to microbial model systems comprising three genetically distinct strains of \emph{Escherichia coli} which cyclically dominate each other like in the children's game rock-paper-scissors~\cite{kerr-2002-418,Kerr}.

These experimental studies of microbial model systems  have motivated a large body of theoretical literature exploring the role of cyclic interactions in ecological systems~
\cite{%LV well-mixed
Itoh1971,Frean2001,Reichenbach2006,Claussen2008,Cremer2008,Berr2009,Andrae2010,Mobilia2010,Dobrinevski2012,Traulsen2012,
%ML well-mixed
May1975,
%LV spatial
Tainaka1988,Tainaka1989,Tainaka1994,frachebourg-1996-54, Frachebourg:1996p5895,Durrett:1998p203,Czaran2002,Reichenbach2008b,He2010,Winkler2010,Venkat2010,Ni2010,Frey2010,He2012,Roman2012,
%ML spatial
Reichenbach2007a,Reichenbach2007,Reichenbach2008b,Peltomaki2008,Reichenbach2008,Muller2010,Ni2010,Rulands2011,He2011,Jiang2011,Frey2010,Lamouroux2012a,Avelino2012}.
Most of this work has focused on two paradigmatic examples of three-species models with cyclic interactions. In a first class of models, direct competition between two individuals leads to the immediate replacement of the weaker species by the stronger one~\cite{Reichenbach2006,Claussen2008,Cremer2008,Berr2009,Andrae2010,Mobilia2010,Dobrinevski2012,Traulsen2012,Tainaka1988,Tainaka1989,Tainaka1994,frachebourg-1996-54, Frachebourg:1996p5895,Durrett:1998p203,Czaran2002,Reichenbach2008b,He2010,Winkler2010,Venkat2010,Ni2010,He2012,Roman2012}. This type of competition, where selection and reproduction are combined into a single process, is similar as in the classical two-species Lotka-Volterra model~\cite{Lotka1920,volterra-1926-31,Dobramysl2008}. The interaction scheme of this cyclic Lotka-Volterra model may be summarized by a set of chemical reactions between the three species $A$, $B$, and $C$:
\begin{eqnarray}
AB\stackrel{}{\rightarrow}AA,\; & BC\stackrel{}{\rightarrow}BB,\; & CA\stackrel
{}{\rightarrow}CC\, .
\label{eq:rr_lv}
\end{eqnarray}
In the second class of models, originally proposed by May and Leonard~\cite{May1975}, selection and reproduction are two separate processes. An interaction between two individuals with different traits (strategies) leads to the death of the weaker species and thereby to empty spaces. Reproduction then follows as a second process which recolonizes this empty space. The ensuing reaction scheme reads:
\begin{subequations}
\begin{align}
AB\stackrel{}{\rightarrow}A\emptyset\,,\; \qquad
& BC\stackrel{}{\rightarrow}B\emptyset\,,\; 
& CA\stackrel{}{\rightarrow}C\emptyset\, ,\\
A\emptyset\stackrel{}{\rightarrow}AA\,,\; \qquad
& B\emptyset\stackrel{}{\rightarrow}BB\,,\; 
& C\emptyset\stackrel
{}{\rightarrow}CC\, .
\end{align}\end{subequations}

Both of these models exhibit absorbing states where all but one species have died out. Due to the inevitable demographic fluctuations in systems with a finite number of individuals these absorbing states will with certainty be reached if one just waits long enough. How long one has to wait strongly depends on the type of model and the ecological scenario under consideration. 

In well-mixed systems, the typical extinction time $T$ was found to scale linearly  with the population size $N$ for the cyclic Lotka-Volterra model~\cite{Claussen2008,Dobrinevski2012,boland-2009-79,parker-2009-80,Reichenbach2006} and  logarithmically for the May-Leonard model~\cite{Reichenbach2007}. The reason for the difference is the different nature of the phase space orbits characterizing the nonlinear dynamics of these two models~\cite{Frey2010,Frey2010}. While the phase portrait of the cyclic Lotka-Volterra model exhibits neutrally stable orbits, the May-Leonard model is characterized by heteroclinic orbits emerging from orbits which spiral out from an unstable reactive fixed point. For neutrally stable orbits, the stochastic dynamics performs an unbiased random walk which implies that $T \propto N$. In contrast, unstable orbits generate a drift of the trajectories in phase space towards the boundary such that the extinction process towards the absorbing states is exponentially accelerated with $T \propto \ln N$~\cite{Frey2010,Mobilia2010,Knebel2013}.

In spatially extended systems, the scaling of $T$ with population size strongly depends on the degree of mixing. In particular, it has been shown for both models that there exists a mobility threshold below which extinction times scale exponentially in the system size. For the May-Leonard model this has been attributed to the existence of spiral waves, which emerge as a result of the local nature of reactions and internal noise~\cite{Reichenbach2007,Reichenbach2007a,Reichenbach2008}. Above a certain mobility the characteristic wave length of the spirals exceeds the system size, effectively rendering the dynamics well-mixed. In this regime, extinctions occurs rapidly. In the cyclic Lotka-Volterra model, spatial patterns are unstable as a result of an Eckhaus instability~\cite{Reichenbach2008b}. However, below a mobility threshold biodiversity is still maintained by strong spatial correlations. Further work has extended these findings to asymmetric reaction rates~\cite{Venkat2010, Berr2009} and more complex interaction networks~\cite{Szabo2007a, Szabo2008, Roman2012,Knebel2013}. In a niche model it has been shown that interaction networks with a high connectivity and a hierarchical or cyclic interaction structure lead to increased diversity~\cite{Mitarai2012,Mathiesen2011}. For the May-Leonard and the cyclic Lotka-Volterra model it was found that spatially inhomogeneous reaction rates have only minor effects on the dynamics~\cite{He2011, He2010}. For the classical two-species Lotka-Volterra model, analytical studies have been performed to understand the underlying mechanism leading to the stabilizing correlations~\cite{Abta2007, Abta2007a}. These studies argue that the stabilization can by understood by the desynchronization of diffusively coupled oscillators. The desynchronization is a result of the combined effect of noise, migration and the dependence of the oscillations' frequency upon their amplitude.

For the one-dimensional May-Leonard model the dynamics leading to extinction has been studied in greater detail. If the individuals diffuse only little or do not diffuse at all, coarse-graining of species' domains has been identified as the dominant dynamical process leading to extinction~\cite{Tainaka1988, frachebourg-1996-54, Frachebourg:1996p5895, He2010}. With increasing diffusion constant other types of collective excitations become important~\cite{Rulands2011}. The dynamics to extinction is then surprisingly rich, comprising rapid extinction, global oscillatory behavior, and traveling waves. The latter involve oscillating overall densities, \emph{i.e.} the domain sizes for the different species change periodically. The statistical weights of these dynamical regimes change qualitatively at threshold values of the mobility and the system size. Taken together, it has turned out that the dynamics in the one-dimensional May-Leonard model is highly complex, much more than one would naively anticipate.

In this manuscript we extend these studies to two-dimensional models with cyclic competition between species. Specifically, we study a generic model comprising both direct competition between species as in the cyclic Lotka-Volterra model, and separated selection and reproduction processes as in the May-Leonard model. Our goal is to identify and characterize the dynamic processes which are responsible for the transient maintenance of biodiversity and which finally lead to the extinction of all but one species. In particular, we are interested in how factors like species mobility and the relative strength of the different competition types govern the complex spatio-temporal dynamics of the system. Employing extensive numerical simulations, we show that the dominant dynamic processes responsible for the transient maintenance of biodiversity correspond to attractors of the global dynamics, \emph{i.e.} the overall density of species in the system. The characteristic features of these attractors give information about the scaling of extinction times with the system size and thereby the stability of biodiversity. Importantly, the attractors change qualitatively at certain threshold values of the mobility and the relative strength of the different competition types. The phase transitions at these threshold values correspond to abrupt changes of the scaling of the extinction time $T$ with population size $N$. These global attractors can be envisioned as minima in an effective free energy landscape. As their counterparts from equilibrium thermodynamics, they give valuable information about the physics underlying the observed transitions and thereby give insight into the mechanisms leading to the stability of ecosystems. Our numerical studies are complemented by scaling arguments based on properties of the complex Ginzburg-Landau equation~\cite{Aranson2002,Reichenbach2008b,Reichenbach2007,Reichenbach2008,Reichenbach2007a}. 

\section{A generic model of cyclically interacting species\label{sec:model}}

We consider a spatially extended population consisting of three distinct species $A$, $B$ and $C$ that compete with each other cyclically in two different ways: either by immediately replacing the competitor
by an individual of its own kind, or by killing the inferior species and creating an empty site $\emptyset$. In addition, individuals may also reproduce if empty spaces are available. These processes are summarized by the following reaction scheme:
\begin{subequations}
\label{eq:rr}
\begin{align}
& AB\stackrel{\sigma}{\rightarrow} A\emptyset,
& A\emptyset\stackrel{\mu}{\rightarrow} AA, \qquad
& AB\stackrel{\nu}{\rightarrow}AA,
\label{eq:rra} \\
& BC\stackrel{\sigma}{\rightarrow} B\emptyset, 
& B\emptyset\stackrel{\mu}{\rightarrow} BB, \qquad
& BC\stackrel{\nu}{\rightarrow}BB, 
\label{eq:rrb} \\
&CA\stackrel{\sigma}{\rightarrow} C\emptyset, 
& C\emptyset\stackrel{\mu}{\rightarrow} CC,  \qquad
& CA\stackrel{\nu}{\rightarrow}CC.
\label{eq:rrc}
\end{align}
\end{subequations}
The reaction rules (\ref{eq:rr}) describe two competing types of selection processes: On the one hand, with rates $\sigma$ and $\mu$, selection and reproduction are separate processes. Selection produces empty sites which are in turn required for reproduction. An empty space is not necessarily occupied by the individual who produced it. We refer to these processes as May-Leonard processes. On the other hand, Lotka-Volterra processes, with a rate $\nu$,  couple selection and reproduction: success in competition directly translates into reproduction. In the following, when we use the term Lotka-Volterra process, this will always imply that the reactions are cyclic. There are two limiting cases which correspond to well established models:  for $\nu\rightarrow 0$ and for $\mu=\sigma=0$ we recover the May-Leonard model \cite{May1975} and a three species model with cyclic interactions of Lotka-Volterra type \cite{Lotka1920,volterra-1926-31, Itoh1971}, respectively. 

\subsection{Stochastic lattice gas model}

We consider a two-dimensional square lattice and employ periodic boundary conditions~\footnote{We also performed simulations on hexagonal lattices and found identical results. To investigate the influence of the ``macroscopic'' lattice structure we performed simulations on hexagonal lattices with hexagonal boundaries and corresponding periodic boundary conditions. While the details of some spatio-temporal patterns and the specific critical values changed slightly, we observed a very similar behavior of the global dynamics and the observed attractors. Note, however, that planar waves and the pairwise appearance of spirals are not observed for absorbing boundary conditions.}. The linear dimension of the lattice is taken as the basic length unit such that the lattice constant $a=1/L$ with $L$ the number of lattice sites along each axis. At each site a fixed number $M$ of individuals ($A$, $B$, $C$ or empty spaces $\emptyset$) are located. $M$ may be viewed as the carrying capacity of a lattice site.  In addition, individuals are also able to move on the lattice.  While the reactions, Eqs.~\eqref{eq:rra}-\eqref{eq:rrc}, are assumed to occur on the same lattice site, the individuals' mobility is implemented as an exchange process at a rate $\epsilon$ between neighboring sites, $XY\stackrel{\epsilon}{\rightarrow}YX$, where $X$ and $Y$ denote species $A$, $B$ and $C$ or empty spaces $\emptyset$. Macroscopically the nearest neighbor exchange process leads to diffusion with an effective diffusion constant $D=\epsilon L^{-2}/2$~\cite{Reichenbach2008,Reichenbach2007,Reichenbach2007a}. \changed{As two particles are involved in migration, it also induces some additional nonlinear reaction terms, which we neglect here~\cite{Lugo2008,Butler2009}.}

\begin{figure}[htb]
\begin{center}
\includegraphics[width=0.9\columnwidth]{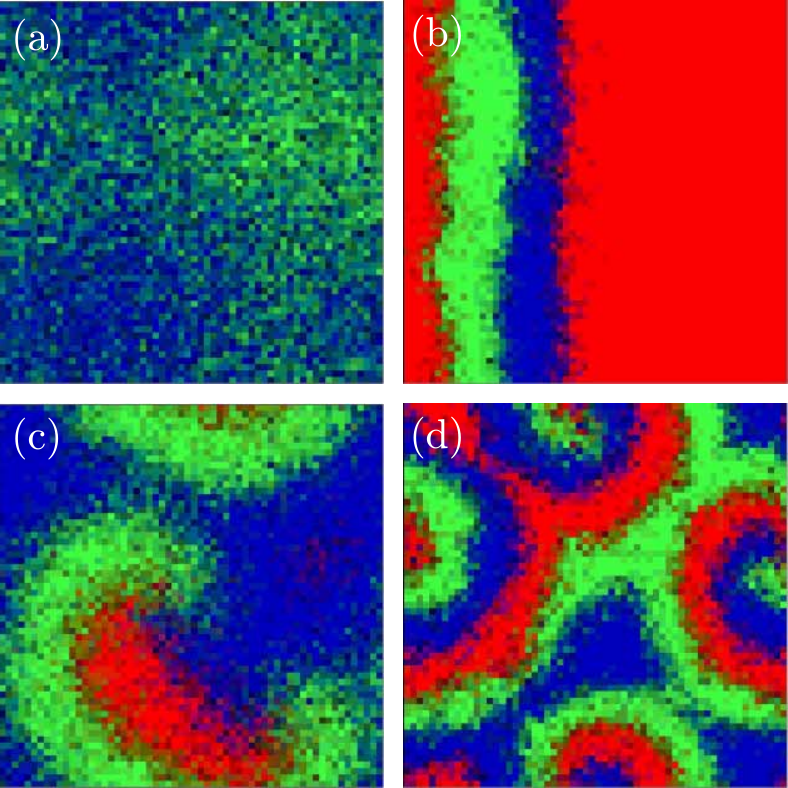}
\caption{(Color online) Typical spatial patterns in the May-Leonard model. Color (gray scale) denotes the concentration of species $A$, $B$ and $C$, with red (medium gray) signifying a site dominated by species $A$, green (light gray) a site dominated by species $B$, and blue (dark gray) being a site dominated by species $C$. (a) For large diffusion constants ($D=1.5 \cdot 10^{-3}$), the dynamics shows global oscillations with periodic switching between states with mainly one species present. (b) For intermediate values of the diffusion constant ($D=5 \cdot 10^{-4}$), we observe planar traveling waves. Here, two of the domains take a characteristic domain size dictated by the diffusion constant. The third domain then occupies the rest of the system.  (c), (d) For even smaller diffusion constants  ($D=3 \cdot 10^{-4}$, and $D=6 \cdot 10^{-5}$),  pairs of rotating spirals appear. The vertices of these spirals move very slowly on a time scale much larger than the time scale corresponding to the propagation speed of their arms. The wave length of the spirals decreases when the diffusion constant is reduced. The system size for all snapshots was $L=60$, and the carrying capacity for each site was chosen to be $M=8$.}\label{fig:snapshots_ml}
\end{center}
\end{figure}

We performed extensive simulations of the ensuing stochastic particle dynamics employing a sequential updating algorithm: At each simulation step an individual is chosen at random. It then either reacts with another also randomly chosen individual from the same site, or is exchanged with an individual of a neighboring stack; each stochastic event occurs with probabilities corresponding to the respective reaction rates $\mu$, $\sigma$, $\nu$ and $\epsilon$. Typical snapshots of the stochastic simulations for the May-Leonard model are shown in Fig.~\ref{fig:snapshots_ml}.

The effective size of the system is $N=M\cdot L^2$. If $M$ and $L$ are large enough, the strength of fluctuations is proportional to $1/\sqrt{N}$ \cite{Gardiner}. The simulation results therefore do not depend on the specific choice of $M$ and $L$, as long as both are not too small and the net system size is kept constant. In particular, the lattice spacing $a=L^{-1}$ must be much smaller than the correlation length $\xi$. 

\changed{Different reaction rates for the species should not limit the validity of our results, as long as the differences between the species are not too large. It has recently been shown that small asymmetries in the reaction rates do not alter the dynamics~\cite{Peltomaki2008}. A general discussion of species-dependent reactions rates is given in Refs.~\cite{He2010,Venkat2010}. In the following we will also set $\mu=\sigma$. While the relation between the selection and reproduction rates in the May-Leonard model affects certain properties of the dynamics (such as the wavelength and velocity of spiral waves), qualitatively the results remain the same~\cite{Reichenbach2008}. This view is supported by some sample runs that we carried out for different values of $\mu/\sigma$.  It is, however, important to note that our results are not valid for extreme choices of the rate, corresponding, for example, to two species predator prey models~\cite{Dobramysl2008,Taeuber2012,Dobramysl2013}.  In all simulations the initial condition was chosen to be a randomly populated lattice with average concentrations corresponding to the reactive fixed point of the well mixed model.}

We fixed the time scale by setting $\mu=\sigma=1$ for the May-Leonard and $\nu=1$ for the Lotka-Volterra limit. The diffusion constant $D$ then gives the mean square displacement of an average particle between two reactions. As an example, with the system size as the unit length a value of $D=10^{-3}$ implies that a particle covers an area of one thousandth of the system size between two succeeding reactions. \changed{We study a regime, where the correlation length is much larger then the lattice spacing. As a result, the lattice spacing is irrelevant as a length scale. The only relevant quantity is the ratio of the characteristic length of spatial patterns given by $D$ and the system size.}

\subsection{Well-mixed limit and invariant manifolds}
\changed{For large populations, intrinsic fluctuations are negligible in our stochastic lattice gas model.} If in addition every individual can interact with every other individual with equal probability, \emph{i.e.} for a well-mixed system, the dynamics is aptly described by deterministic rate equations for the densities $a$, $b$ and $c$ of the species $A$, $B$ and $C$, respectively:
\begin{subequations}
\label{eq:re}
\begin{align}
\partial_t a &= -\sigma a c +\mu a(1-\rho) +\nu a (b-c), \\
\partial_t b &= -\sigma b a +\mu b(1-\rho) +\nu b (c-a),\\
\partial_t c &= -\sigma c b +\mu c(1-\rho) +\nu c (a-b). 
\end{align}
\end{subequations}
Here, $\rho = a + b + c$ is the total density of species and $0\leq a,b,c,\rho\leq 1$. Equation~\eqref{eq:re} exhibits three absorbing fixed points: $(1,0,0)$, $(0, 1, 0)$, and $(0,0,1)$. They correspond to the extinction of two of the three species. Another fixed point at $(0,0,0)$ corresponds to the extinction of all three species. However, this fixed point cannot be reached by the stochastic dynamics from the initial conditions we study here. In addition, there is a reactive fixed point 
\begin{eqnarray}
(a^*,b^*,c^*)=\frac{\mu}{3\mu +\sigma}(1,1,1) ,
\end{eqnarray}
at which all three species coexist. The dynamics in the vicinity of the reactive fixed point can be studied by linearizing Eq.~(\ref{eq:re}) around $(a^*,b^*,c^*)$ and by determining the eigenvalues of the corresponding Jacobian. We find that the dynamics close to the reactive fixed point is characterized by an attractive eigendirection with a negative eigenvalue $\kappa_0=-\mu$ and two further eigendirections with eigenvalues 
\begin{equation}
\kappa_{\pm}=\frac{1}{2}\frac{\mu}{3\mu+\sigma}\left[ \left(1\pm \mathrm{i} \sqrt{3}\right) \sigma \pm \mathrm{i} \, 2 \sqrt{3} \nu \right].\label{eq:eigenvalues}\, 
\end{equation}

Therefore, the eigenvectors corresponding to $\kappa_{\pm}$ span, to linear order, an invariant manifold onto which the dynamics relaxes exponentially fast. To obtain an approximation for the invariant manifold, valid to second order in the concentrations, we follow the steps given in Ref.~\cite{Reichenbach2008}. We first transform to a new reference frame whose origin is the unstable fixed point, $(x_A,x_B, x_C)=(a-a^*, b-b^*, c-c^*)$.  Further, we choose the eigendirections of the fixed point as basis vectors for our new reference frame. To this end, we employ a rotation of the coordinate system, 
\begin{equation}
\vec{y}=\frac{1}{3}\left(\begin{array}{ccc}
\sqrt{3} & 0 & -\sqrt{3}\\
-1 & 2 & -1\label{eq:ycoord}\\
1 & 1 & 1\end{array}\right)\vec{x}\, .
\end{equation}
The stable eigendirection corresponding to $\kappa_0$ is then given by the $y_C$-direction, while $y_A$ and $y_B$ span the invariant manifold to linear order. We parametrize the invariant manifold by $y_{C}=G(y_{A},y_{B})$.  Using the ansatz $G(y_{A},y_{B})\sim y_{A}^{2}+y_{B}^{2}$
and determining the proportionality constant such that 
\begin{equation*}
\partial_{t}G(y_{A}(t),y_{B}(t))=\partial_{y_{A}}G\cdot\partial_{t}y_{A}+\partial_{y_{B}}G\cdot\partial_{t}y_{B}\stackrel{!}{=}\partial_{t}y_{C}\big|_{y_{C}=G}
\end{equation*} 
we find that $G(y_A,y_B)$ is, to second order, given by
\begin{equation}
G(y_{A},y_{B})=\frac{\sigma}{4\mu}\frac{3\mu+\sigma}{3\mu+2\sigma}\left(y_{A}^{2}+y_{B}^{2}\right).\,\label{eq:invariantmanifold}\end{equation}
This equation is valid only for $\mu\neq0$. In the limit of a cyclic Lotka-Volterra model, $\mu=0=\sigma$, we find $G_{\mu=0,\sigma=0}(y_{A},y_{B})=0$ and the invariant manifold is given by the unit simplex defined by $a+b+c=1$. This result can also be directly inferred from the Lotka-Volterra reactions, which preserve the total density and thereby lead to dynamics on an invariant manifold given by $a(t)+b(t)+c(t)=1$.

The rate equations in the new reference frame read:
\begin{subequations}
\begin{align}
\partial_{t}y_{A} & = \frac{\sqrt{3}}{4} (2 \nu +\sigma ) \left(y_A^2-y_B^2\right)\\
&+\frac{\sqrt{3}}{2}  (2 \nu +\sigma ) y_B \left(\frac{\mu }{3 \mu +\sigma }+y_C\right)\nonumber\\
&+\frac{y_A \left\{\mu  \sigma -(3 \mu +\sigma ) \left[\sigma  y_B+(6 \mu +\sigma ) y_C\right]\right\}}{2 (3 \mu +\sigma )}\nonumber\\
&+\frac{\sqrt{3} (2 \nu +\sigma ) y_B \left[\mu +(3 \mu +\sigma ) y_C\right]}{2 (3 \mu +\sigma )}\,,\nonumber\\
\partial_{t}y_{B} & = -\frac{1}{4} \sigma ^2( y_A^2- y_B^2)\\
&-\frac{\sqrt{3}(2\nu+\sigma)}{2} y_A \left(\frac{ \mu  }{ 3 \mu +\sigma }+  y_B +  y_C\right)\nonumber\\
&+\frac{y_B}{2(3\mu +\sigma)} \left[\mu  \sigma -(3 \mu +\sigma ) (6 \mu +\sigma ) y_C\right],\nonumber \\
\partial_{t}y_{C} & = -\mu y_C -(3\mu +\sigma) y_C^2 +\frac{\sigma}{4}(y_A^2+y_B^2).
\label{eq:yevolution}
\end{align}
\end{subequations}
What is the simplest differential equation that captures the essential features of the rate equations \eqref{eq:re}? Such a differential equation is called normal form, and is obtained by a nonlinear transformation which eliminates the quadratic terms. Following the steps in Ref.~\cite{Reichenbach2008}, one makes a quadratic ansatz for the transformation and determines the coefficients canceling the quadratic terms. We find that the transformation is given by
\begin{subequations}
\begin{align}
z_{A} & =  y_{A}+\alpha_{1}\left(\sqrt{3}y_{A}^{2}+\alpha_{2}y_{A}y_{B}-\sqrt{3}y_{B}^{2}\right)\, , \\
z_{B} & =  y_{B}+\alpha_{1}\left(\frac{\alpha_{2}}{2}y_{A}^{2}-2\sqrt{3}y_{A}y_{B}-\frac{\alpha_{2}}{2}y_{B}^{2}\right),\,\label{eq:nonlineartrafo}
\end{align}
\end{subequations}
with prefactors
\begin{subequations}
\begin{align}
\alpha_{1} &=  \frac{3\mu+\sigma}{28\mu}\frac{7(2\nu+\sigma)\sigma}{27\nu^{2}+27\nu\sigma+7\sigma^{2}}\, ,\\
\alpha_{2} &=  10+\frac{18\nu}{\sigma}-\frac{2\nu}{2\nu+\sigma}\, .
\end{align}
\end{subequations}
Upon introducing a complex  variable $z = z_A+\text{i} z_B$ and neglecting terms of order $\mathcal{O}(z^4)$, the dynamics can finally be written in the form 
\begin{equation}
\partial_t z =  (c_1 -\mathrm{i}\omega)z - c_2(1-\mathrm{i}c_3)|z|^2 z\, ,\label{eq:cgle}
\end{equation}
where
\begin{subequations}
\begin{align}
\omega & =  \frac{\sqrt{3}}{2}\frac{\mu(2\nu+\sigma)}{3\mu+\sigma}\, ,\\
c_{1} & =  \frac{1}{2}\frac{\mu\sigma}{3\mu+\sigma}\, , \\
c_{2} & =  \frac{\sigma(3\mu+\sigma)}{56\mu(3\mu+2\sigma)}\\
&\times\frac{\sigma^{2}(48\mu+11\sigma)+3\nu(60\mu+13\sigma)(\nu+\sigma)}{\sigma^{2}+\frac{27}{7}\nu(\nu+\sigma)}\, ,\nonumber\\
c_{3} & = \frac{1}{c_2} \frac{(3\mu+\sigma)\sqrt{3}(2\nu+\sigma)}{56\mu(3\mu+2\sigma)} \\*
&\times\frac{\sigma^{2}(18\mu+5\sigma)+9\nu(6\mu+\sigma)(\nu+\sigma)}{\sigma^{2}+\frac{27}{7}\nu(\nu+\sigma)}\nonumber\, .
\end{align}
\end{subequations}
While the limiting case of a May-Leonard model is found by simply setting $\nu = 0$, the cyclic Lotka-Volterra model is recovered by first taking the limit $\sigma\rightarrow0$ and then $\mu\rightarrow 0$. Other ways for performing this limit are possible. However, taking the limit in $\sigma$ first, ensures that we obtain the established cyclic Lotka-Volterra model, which does not comprise empty sites: $a+b+c=1$~\footnote{Explicitly, consider taking the limit in the total equilibrium concentration $a^*+b^*+c^*= {1}/({1 + \sigma/3\mu})$. This dictates that in order to obtain a system without empty sites $\sigma/\mu\rightarrow 0$, \emph{i.e.} the production rate of empty sites per concentration must vanish faster than the consumption rate.}.

\subsection{Spatially extended continuum model}

In a continuum formulation, the nearest neighbor exchange process macroscopically leads to diffusion with a diffusion constant $D=\epsilon L^{-2}/2$. The ensuing diffusion-reaction equations are simply obtained from the rate equations \eqref{eq:re} by supplementing them with diffusion terms $D\nabla^2 a$, $D \nabla^2 b$, and $D\nabla^2 c$, respectively \cite{Reichenbach2007a,Reichenbach2008b,Reichenbach2007,Reichenbach2008}. Then, upon applying the above transformations to the diffusion terms one obtains
\begin{equation}
\partial_t z = D\left[ \nabla^2 z- \mathrm{i}\left( \vec{\nabla} z^*\right)^2\right]+\text{reaction terms},
\end{equation}
where we have neglected gradient terms of order $\mathcal{O}((\vec{\nabla} z)^3)$.
We expect that the dynamics is dominated by the long wavelength modes, and therefore only keep the leading order gradient term, leading to normal diffusion in the complex concentration $z$. This finally leads to the complex Ginzburg-Landau equation, 
\begin{equation}
\partial_t z = D\nabla^2 z +(c_1 -\mathrm{i}\omega)z - c_2(1-\mathrm{i}c_3)|z|^2 z\, ,\label{eq:cgle}
\end{equation}
a paradigmatic equation in nonlinear dynamics~\cite{Aranson2002}.

\section{The May-Leonard Limit \label{sec:ml}}
\begin{figure*}[htb]
\begin{center}
\includegraphics[width=0.99\textwidth]{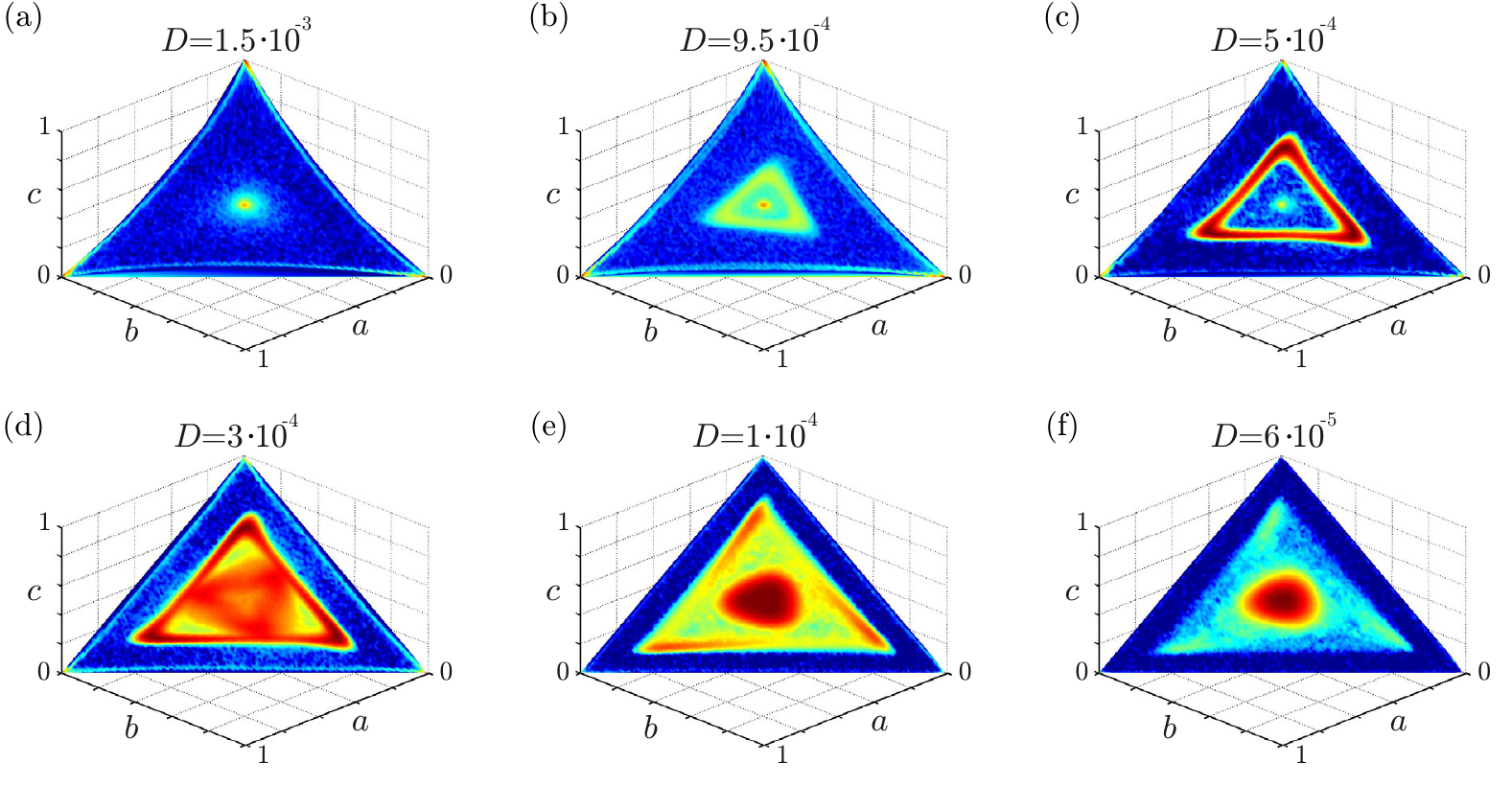}
\caption{(Color online) Free energy landscapes of the May-Leonard model for various values of the diffusion constant $D$. The dynamics of the overall densities $\overline{a}$, $\overline{b}$, $\overline{c}$ is strongly confined to the invariant manifold of the well-mixed model, Eq. \eqref{eq:re_ml}. To study the mechanisms underlying the distinct spatio-temporal patterns found in the spatial, stochastic May-Leonard model we projected the probability for the overall densities onto the invariant manifold of the rate equations (\ref{eq:re_ml}) for different values of the diffusion constant $D$. Color (gray scale) denotes the logarithmic probability to find the system globally in a specific state before reaching one of the absorbing states, such that red (medium gray) denotes a high probability, yellow (light gray) a medium probability and blue (dark gray) a low probability. The absorbing states themselves are not part of the statistics. For large $D$, no stable spatial structures can form and the dynamics corresponds to the well mixed case, Eqs.~\eqref{eq:re_ml}. As $D$ becomes smaller than $D\approx 9.5\cdot 10^{-4} $ an attractor of the global dynamics emerges, effectively stabilizing the system against extinction. This attractor corresponds to planar, traveling waves with oscillating overall densities, and grows in radius with decreasing $D$ due to a decreasing wavelength of the planar waves. For $D\leq 3\cdot 10^{-4}$, a second attractor emerges, corresponding to rotating spirals. As a result of a decreasing wavelength of the spiral patterns the second attractor's radius decreases with the diffusion constant, while the attractor corresponding to the traveling waves diminishes. Parameters were $L=60$ and $M=8$. Each plot was averaged over at least $100$ realizations of the stochastic spatial dynamics.}\label{fig:manifolds_ml}
\end{center}
\end{figure*}
The May-Leonard model, obtained in the limit $\nu\rightarrow 0$, is characterized by the following reduced set of reaction rules:
\begin{subequations}
\label{eq:rr_ml}
\begin{align}
AB\stackrel{\sigma}{\rightarrow}A\emptyset\,,\; \qquad
& BC\stackrel{\sigma}{\rightarrow}B\emptyset\,,\; 
& CA\stackrel
{\sigma}{\rightarrow}C\emptyset\, ,
\label{eq:rr_ml_a} \\
A\emptyset\stackrel{\mu}{\rightarrow}AA\,,\; \qquad
& B\emptyset\stackrel{\mu}{\rightarrow}BB\,,\; 
& C\emptyset\stackrel
{\mu}{\rightarrow}CC\, .
\label{eq:rr_ml_b}
\end{align}
\end{subequations}
For large systems in the well-mixed limit, the dynamics is described by the May-Leonard equations \cite{May1975},
\begin{subequations}
\label{eq:re_ml}
\begin{align}
\partial_t a &= -\sigma a c +\mu a(1-\rho)\, , \label{eq:re_ml_a}\\
\partial_t b &= -\sigma b a +\mu b(1-\rho)\, , \label{eq:re_ml_b}\\
\partial_t c &= -\sigma c b +\mu c(1-\rho)\, . \label{eq:re_ml_c}
\end{align}
\end{subequations}
The nonlinear dynamics of these equations is characterized by the same types of fixed points and invariant manifold as the general model \eqref{eq:re}. The reactive fixed point $(a^*,b^*, c^*)$ is globally unstable, as manifested by the existence of the Lyapunov function  $\mathcal{L}=abc/\rho^3$.  When starting in the vicinity of the unstable fixed point, the trajectories spiral outward on the invariant manifold and form heteroclinic cycles, approaching the boundary of the phase space and the absorbing states without ever reaching them \cite{May1975}. However, intrinsic noise due to the stochastic nature of the interactions, and spatial structure drastically alter the observed behavior. While in well-mixed systems stochastic fluctuations drive the system into one of the absorbing states within a short time proportional to the logarithm of the system size \cite{Reichenbach2006, Dobrinevski2012, cremer-2009-11, Andrae2010, Berr2009, Claussen2008, Traulsen2012}, spatial structures may effectively delay extinction by orders of magnitude~\cite{Reichenbach2007, Durrett:1998p203}.

Similar to the previously studied one-dimensional case~\cite{Rulands2011}, the two-dimensional, stochastic May-Leonard model exhibits distinct  dynamical regimes as a function of the diffusion constant $D$ (Fig.~\ref{fig:snapshots_ml}). From our simulations we find the following phenomenology: For large values of $D$, we observe that the system (after some initial transient) is first almost entirely taken over by one species, but with a few individuals of a second species surviving, which dominates over the more abundant species. This second species will then slowly take over the system and thereby lead to a dynamical behavior that is reminiscent of the heteroclinic orbits of the deterministic, well-mixed system, where the global dynamics approaches the boundary of the invariant manifold. In this regime, spatial patterns are of minor importance and the dynamics can be understood in terms of a quadratic coagulation process as outlined in Ref. \cite{Rulands2011}. With decreasing diffusion constant we observe planar waves of cyclically aligned uniform domains as well as rotating spiral waves [Fig.~\ref{fig:snapshots_ml}(b)-(d)]. In planar waves the overall concentrations may be constant, corresponding to stable domain borders, or change periodically, as a result of ``tunneling" events in the leading edges of the fronts. The leading edges of the fronts may reach into second next domains, \emph{i.e.} there is a finite probability for particles to penetrate domains of prey via ``tunneling'' events~\cite{Rulands2011}. As a consequence domain sizes oscillate periodically between characteristic length scales, thereby leading to oscillating overall densities. The dynamics of rotating spirals has been extensively studied in Refs.~\cite{Reichenbach2007,Peltomaki2008}. Both planar waves and rotating spirals are only metastable, as stochastic fluctuations eventually lead to the annihilation of neighboring fronts and the dynamics will ultimately end in one of the three absorbing states which correspond to the extinction of two of the three species. The dynamics into the absorbing states has been found to be highly nontrivial, as the dynamical regimes described above lead to transitions into the absorbing states on different time scales.  Furthermore, their statistical weight heavily depends on the diffusion coefficient $D$~\cite{Rulands2011,Ni2010,Lamouroux2012a,Reichenbach2007,Reichenbach2007a}.

\subsection{Global attractors and ``free energy landscape'' of the spatio-temporal dynamics}

To gain insight into the mechanisms responsible for these qualitatively different spatio-temporal patterns and how they determine the longevity of biodiversity in the population, we studied the global phase portrait of the dynamics. Figure~\ref{fig:manifolds_ml} shows histograms for the overall concentrations 
\begin{equation}
\bigl( \overline{a} (t), \overline{b} (t), \overline{c} (t) \bigr) 
=\int \bigl( a( {\bf x},t), b( {\bf x},t), c( {\bf x},t) \bigr) \upd^2 x
\end{equation} 
of the three species on the invariant manifold of the rate equations, Eqs.~\eqref{eq:re_ml_a}-\eqref{eq:re_ml_c}.  In detail, the negative logarithm of the probability $P(\overline{a},\overline{b},\overline{c})$ to find the system in a specific global state $(\overline{a},\overline{b},\overline{c})$ before reaching one of the absorbing states is projected onto the invariant manifold: 
\begin{equation}
\mathcal{F}(\overline{a},\overline{b},\overline{c}) \equiv -\ln P(\overline{a},\overline{b},\overline{c}), \label{eq:fe}
\end{equation}
\changed{The quantity $\mathcal{F}$ hence gives the logarithmic density of global trajectories in phase space and it can be considered as an effective potential in the following sense:  When instead of the mean-field reaction term, as given by the right hand side of Eq.~\eqref{eq:cgle}, one uses $\mathcal{F}$ as a ``renormalized" potential for a Ginzburg-Landau theory one obtains a reaction-diffusion equation which gives a good description of the spatio-temporal dynamics. Long-lived spatio-temporal patterns correspond to regions of high probability on the manifold, and are termed \emph{global attractors} of the spatio-temporal dynamics in the following. Equivalently, adapting terminology from statistical mechanics, these attractors may be viewed as minima of the ``free energy landscape'' $\mathcal{F}$. Of course, it is to be understood that these attractors are only metastable, \emph{i.e.} while the system spends a long time in these states, ultimately demographic fluctuations will drive the system into one of the absorbing states. Intuitively, one may visualize those fluctuations as driving the escape of the dynamics from the minima in the ``free energy landscape'' into one of the absorbing states. We will later see that \emph{qualitative} changes in the shape of these minima correspond to transitions in the nature of the dynamic processes leading to extinction and the mean times to extinction. These transitions should not be confused with non-equilibrium phase transitions. Rather they are to be considered as bifurcations in the nonlinear dynamics.}

Figure \ref{fig:manifolds_ml} shows that the shape of these global attractors strongly depends on the magnitude of the diffusion constant $D$. For $D>10^{-3}$, there are no attractors other than the regions in the immediate vicinity of the three absorbing states. All trajectories describing the global dynamics quickly leave the unstable fixed point $(a^*,b^*,c^*)$ and approach the boundaries of the invariant manifold. Therefore, the probability is highest in the center (because the dynamics starts there) and at the boundaries.  In this regime, the system can be considered as well-mixed. The heteroclinic orbits in the global phase portrait then correspond to spatially uniform oscillations between states where one of the three species dominates; cf. Fig.~\ref{fig:snapshots_ml}(a). With decreasing diffusion constant $D$ the nature of the global attractor changes qualitatively. Starting at the center of the manifold, the free energy develops a distinct local minimum which then evolves into a triangular shaped closed region; see  Fig.~\ref{fig:manifolds_ml}(a)-(c). In other words, the phase portrait of the global population dynamics changes from an unstable fixed point with heteroclinic orbits to a pronounced \emph{limit cylce}. Inspecting the spatio-temporal  patterns as obtained from our stochastic simulations, we find that this limit cycle of the global dynamics  corresponds to planar traveling waves; see also Fig.~\ref{fig:snapshots_ml}(b). The triangular shape is the result of oscillating overall densities. In the following we will refer to this particular limit cycle as the \emph{wave attractor}. Further lowering the diffusion constant, a second global attractor emerges as a smaller triangle inside the triangle corresponding to the planar waves, cf.  Fig.~\ref{fig:manifolds_ml}(d)-(f). The inner triangular attractor corresponds to rotating spirals and will henceforth be denotes as the \emph{spiral attractor}. We find that in this regime of diffusion constants the two attractors coexist, meaning that we observe both planar waves and rotating spirals. Both processes may even be found within the same realization. With even further decreasing the diffusion constant, the weight of the triangular-shaped attractor corresponding to planar waves decreases and the attractor eventually disappears completely. As a consequence, the attractor of spiral waves gains weight, such that the dynamics at low mobilities is dominated by spiral waves. Taken together, we find that the phase portrait of the global dynamics changes qualitatively upon decreasing the diffusion constant, and that those qualitative changes have a one-to-one correspondence with distinct spatio-temporal patterns in the population dynamics. As a consequence, the free energy landscape on the invariant manifold can be taken as a fingerprint of the spatio-temporal dynamics. We will use it in the following to identify transitions between different patterns and analyze the ensuing changes in the dependence of the extinction times on system size. 

\subsection{Pattern selection and extinction times}

\begin{figure}[t!]
\begin{center}
\includegraphics[width=0.99\columnwidth]{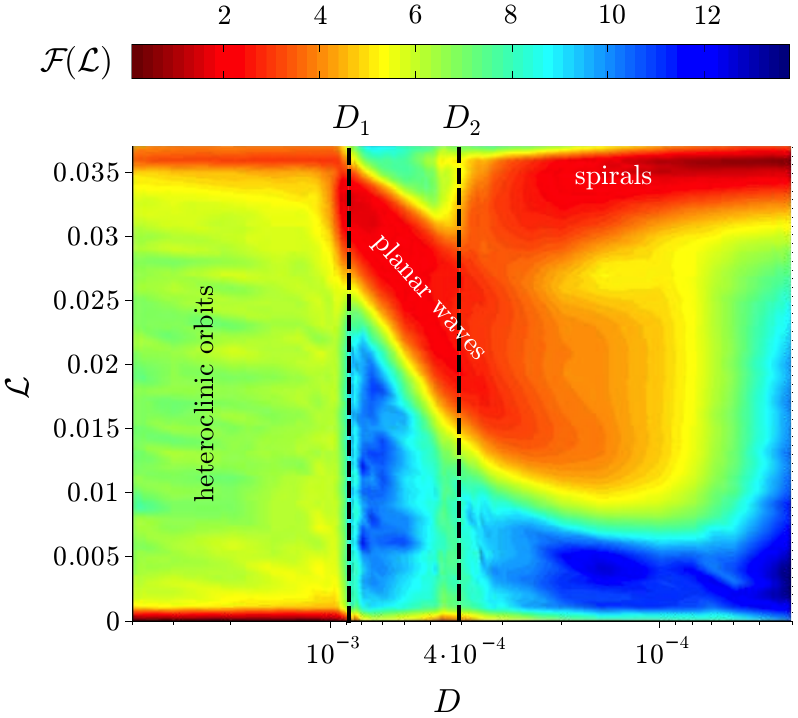}\\
\caption{(Color online) Free energy landscape of the global dynamics of the two-dimensional May-Leonard system. The value of the Lyapunov function $\mathcal{L}$ is a measure of how close a specific state is to the boundary of the invariant manifold ($\mathcal{L}=0$). Color (gray scale) denotes the logarithmic probability to find the system at a specific value of $\mathcal{L}$, \emph{i.e.} the free energy formally defined in Eq.~\eqref{eq:fe}. Red (medium gray) signifies small values of the free energy (minima of the potential) and thereby an attractor of the global dynamics. Yellow (light gray) denotes intermediate values, and blue large values of the free energy. The free energy landscape changes qualitatively at two threshold values for the diffusion constant $D$. For large values of $D$ the effective free energy has minima in the center ($\mathcal{L}=0.037$), where the dynamics starts, and at the boundaries of the invariant manifold ($\mathcal{L}=0$). At a first threshold $D_1\approx 9\cdot 10^{-4}$ an attractor emerges, which moves away from the reactive fixed point ($\mathcal{L}=0.037$) with decreasing values of $D$. Below a second threshold, $D\approx 4.5\cdot 10^{-4}$, a second attractor emerges near the reactive fixed point, coexisting with the first one. For even smaller mobilities the dynamics is solely determined by the attractor near the reactive fixed point. Comparing with our simulations we find that these attractors correspond to global oscillations (heteroclinic orbits), planar waves and rotating spirals, respectively.  The stochastic simulations were performed on a square lattice of linear size $L=60$ and with a carrying capacity $M=8$ for each site. For each values of $D$, the histogram was averaged over at least $100$ realizations.}
\label{fig:phaseportrait_ml}
\end{center}
\end{figure}

The attractors in Fig.~\ref{fig:manifolds_ml} show a triangular symmetry. A reduced representation for  the global dynamics on the invariant manifold can therefore be obtained in terms of a properly defined radial variable.  A convenient choice is the Lyapunov function 
\begin{equation}
\mathcal{L}
\equiv \frac{\overline{a}\,\overline{b}\,\overline{c}}{\left( \overline{a}+\overline{b}+\overline{c}\right)^3}
\end{equation} 
evaluated with the global concentrations $\overline{a}$, $\overline{b}$, $\overline{c}$. It measures the distance of a global state to the boundaries of the invariant manifold and is approximately constant along the attractor for the planar waves. Figure~\ref{fig:phaseportrait_ml} shows the effective free energy $\mathcal{F}(\overline{a},\overline{b},\overline{c})$ as a function of the Lyapunov function and the diffusion constant. One easily identifies two threshold values of the diffusion constant where there are qualitative changes in the free energy landscape. We recover a threshold value $D_1 \approx 9\cdot 10^{-4}$ marking a transition from a well-mixed dynamics to a dynamics with spatio-temporal patterns~\cite{Reichenbach2007,Reichenbach2007a}. However, the range of patterns is much richer than previously noted. Actually, the first threshold $D_1$ marks a transition from spatially uniform oscillations between states dominated by a single species to planar waves where the three species cyclically chase each other. Note that the global oscillations still form part of the dynamics, albeit with a lower probability.  Upon lowering the diffusion constant below a second threshold value, $D_2\approx 4.5\cdot 10^{-4}$, the histogram of system trajectories becomes bimodal with a second metastable attractor emerging which is located close to the center of the invariant manifold. It can hence be identified with the inner, triangular attractor on the invariant manifold. As discussed before, this second attractor corresponds to rotating spiral waves. The coexistence of two attractors in this regime of mobilities means that both, planar waves and rotating spirals, are observed. Depending on the choice of initial conditions, the dynamics may at first end up in either one of the two attractors. Due to stochastic fluctuations it may, however, from time to time switch 
between the two attractors akin to thermal fluctuations causing rare transitions between different potential minima. With further decreasing $D$ we observe that the metastable attractor corresponding to planar waves dissolves and only the attractor corresponding to rotating spirals remains.

\begin{figure}[t]
\begin{center}
\includegraphics[width=0.99\columnwidth]{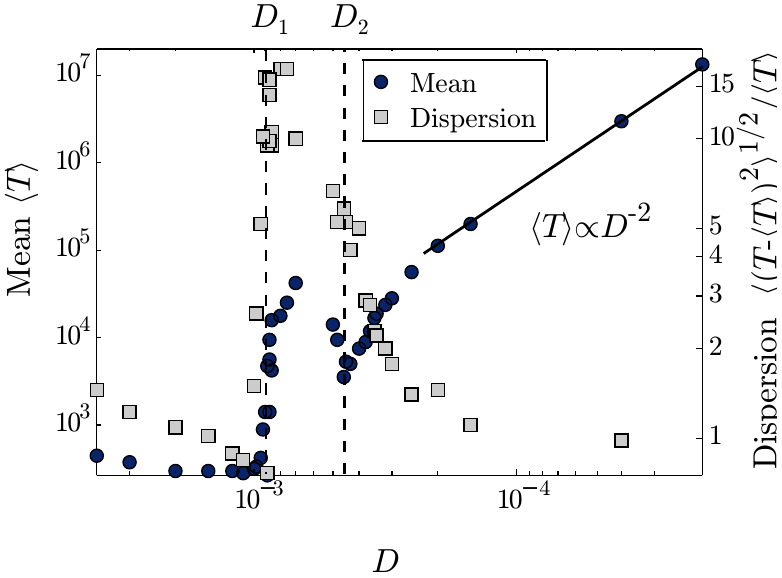}
\caption{Mean lifetimes (dark dots) and coefficient of variation (gray squares) of the two-dimensional May-Leonard system as a function of the diffusion constant $D$. The mean lifetime increases abruptly at a first threshold value $D_1$ (indicated by a dashed line) where planar waves form. After passing through a maximum the lifetime decreases again. This is due to planar waves which become unstable with decreasing correlation length. At the second threshold $D_2$ (dashed line) rotating spirals become possible. With further decreasing diffusion constant the mean lifetime is asymptotically described by a power law $D^{-2}$. The coefficient of variation is a dimensionless measure for the dispersion of the probability distribution of $T$. The dispersion near the upper threshold $D_1$ becomes large, \emph{i.e.} we observe dynamical regimes on a variety of different time scales.}
\label{fig:lifetimes_ml}
\end{center}
\end{figure}

To further scrutinize the effect of these spatio-temporal patterns and the ensuing metastable global attractors on the system's dynamics, we analyzed the mean first passage time into the absorbing states as a function of $D$, see Fig. \ref{fig:lifetimes_ml}. 
We find that the mean time to extinction increases abruptly at $D_1$, where the global attractor of planar waves emerges.  After passing a peak value the mean lifetime then decreases again, as the wavelength of the planar waves becomes smaller. Then, as a result, these waves become more prone to fluctuations, and the rate of domain annihilation increases. Finally, below $D_2$ the lifetime increases again, which we attribute to the emergence of stable spiral waves. For small values of $D$ the mean lifetime follows a power law $\left< T\right>\propto D^{-2}$. This dependence can be understood by a simple scaling argument: Since spirals annihilate pairwise as they meet, the mean lifetime should scale quadratically with their number, $\langle T\rangle \propto (n_{\text{spirals}})^2$. The number of spirals in the system scales with their wavelength as $n_{\text{spirals}}\propto \lambda^{-2}$. With $\lambda\propto\sqrt{D}$ we then infer that the mean lifetime scales as $\langle T\rangle\propto D^{-2}$, which is in good agreement with our numerical results. 

Figure~\ref{fig:lifetimes_ml} also shows the coefficient of variation, defined as the standard deviation divided by the mean, 
\begin{equation}
c_v\equiv\slfrac{\sqrt{\left<\left(T-\left<T\right>\right)^2\right>}}{\left< T\right>}\, .
\end{equation} 
It gives a dimensionless measure for the dispersion of the probability distribution of $T$. We find that the dispersion increases drastically right at the threshold $D_1$. In this regime the standard deviation is much larger than the mean. From the spatio-temporal dynamics observed in our simulations we infer that this is due to the fact that there are several distinct dynamic processes driving the system towards an absorbing state and that these processes occur on greatly different time scales. There are rapid extinction processes, where - after a short transient - domains in a planar wave are aligned in a non-cyclic order and thus immediately annihilate. We also find a process, where the global dynamics performs heteroclinic orbits. Last, one observes metastable planar waves, cf. Fig.~\ref{fig:snapshots_ml}. Note that although the planar waves process is metastable it does not necessary mean that it dominates the long time properties of the system. In Ref.~\cite{Rulands2011} it has been shown for the one dimensional model that probability of extinction scales differently with the system size for these two processes. In particular, one observes a crossover, such that for small systems planar waves determine the long time tails, while for very large systems global heteroclinic orbits are responsible for the longest living states.  The relative weight of these processes depends on the diffusion coefficient $D$. As we have already learned from the above analyses, below the lower threshold value, $D_2$, there are also spiral waves emerging. With decreasing $D$, spirals become the dominant patterns while all the other dynamic processes become less and less probable. As a result, the mean time to extinction is dominated by an escape out of the spiral attractor. The dispersion therefore decreases again.

The probability distributions of first passage times of the above dynamical processes leading into the absorbing states show significantly different scaling behavior with the system size, cf. Ref.~\cite{Rulands2011}. From an evolutionary perspective, the tails of these distributions are most relevant because they correspond to rare, but extremely long-living communities maintaining biodiversity. The reason for their relevance is that the probability to observe a short-living (transient) ecosystem in nature is much lower than the probability to observe an ecosystem which persists for a long time. In Ref.~\cite{Rulands2011} two of the authors showed that the tail of the distributions of first passage times of heteroclinic orbits scale like $\exp\left(T/N\right)$, while for traveling waves the tail scales like $\exp\left[ T/(\ln N)^3 \right]$. As a consequence, there is a crossover in the tail of the overall distribution of first passage times. Interestingly, while for small systems the long time dynamics is dominated by traveling waves, for large systems it is dominated by heteroclinic orbits. Although the computation of the distribution of first passage times is not feasible in two dimensions, we expect that similar arguments will hold here, as well. 

As shown in Ref. \cite{Reichenbach2007, Reichenbach2007a}, there is a transition from a spatially uniform dynamics reminiscent of a well-mixed system to a dynamics dominated by spatio-temporal patterns when the wavelength of the pattern exceeds the system size. Following the classical theory of front propagation into unstable states \cite{Saarlos2003}, the wavelength of the traveling and spiral waves can be determined using the complex Ginzburg-Landau equation (\ref{eq:cgle})~\cite{Reichenbach2008,Reichenbach2007a}
\begin{equation}
\lambda = -\frac{2\pi c_3}{\sqrt{c_1}\left(1-\sqrt{1+c_3^2}\right)}\, \sqrt{D} \, .
\label{eq:wavelength}
\end{equation} 
Due to the difference in geometry between planar and spiral waves this implies two distinct thresholds, 
$D_1$ and $D_2$. For planar waves on a square lattice we simply have the condition that the wavelength equals the system size, $\lambda(D_1) =1$. In Refs.~~\cite{Reichenbach2007, Reichenbach2007a} it was found that the calculated wavelength deviates by a constant factor of 1.6 from the numerical value of the wavelength. \changed{This rescaling factor accounts for the renormalization of the reaction term due to spatio-temporal correlations, as captured by the global attractors.}  Using this rescaling factor we find a threshold value $D_1 \approx 7.6\cdot 10^{-4}$, in good agreement with the numerically found value, $D_1\approx 9\cdot 10^{-4}$, cf. Fig.~\ref{fig:phaseportrait_ml}. The very same threshold is also found in the one dimensional model~\cite{Rulands2011}. There, planar waves are the only possible spatial pattern and the threshold stems from their wavelength outgrowing the system size. Remarkably, the numerical values for $D_1$ coincide in both, one and two spatial dimensions, as the complex Ginzburg-Landau equation predicts equal wavelengths for both cases, see further below. Since spirals always arise as pairs of anti-rotating spirals, stable pairs are possible, as long as the minimum distance $d_{\text{min}}$ between two vertices of the spirals is smaller than half of the system size. In other words, the threshold $D_2$ is given by $d_\text{min}(D_2)=1/2$. To obey geometric constrains dictated by the periodic boundary conditions and the spirals' wavelength, the minimum distance of two anti-rotating spirals is $d_{\text{min}}=2/3 \lambda(D)$, cf. also Fig.~\ref{fig:snapshots_ml}(d). This implies a threshold value of $\lambda=3/4$ which is close to $\lambda\approx 0.8$ obtained numerically in Ref.~\cite{Reichenbach2007}. Hence, from $2/3\lambda(D_2) = 1/2$ we obtain $D_2\approx 4.3\cdot 10^{-4}$, in good agreement with the numerical results shown in Fig.~\ref{fig:phaseportrait_ml}.

In the following we provide a scaling argument giving the scaling of the size of the wave attractor with the mobility $D$. In the intermediate regime between the two threshold values of $D$ the wavelength of the planar wave patterns is of the same order as the system size. Hence, in this regime the finite spatial extension of the system is important. In our case, periodic boundary conditions allow stationary domain profiles only for certain values of the wave length, $\lambda=1/n$, $n=1,2,\ldots$~\footnote{Note that for $n>1$ such patterns are only observed for specific initial conditions.}. If the wavelength does not match any of these values, we observe oscillations in the overall concentrations, corresponding to the triangular attractor in Fig.~\ref{fig:manifolds_ml}. In the intermediate regime, $D_1>D>D_2$, where $\lambda$ is slightly smaller than 1, two domains take the characteristic domain size dictated by $D$ and the third domain occupies the rest of the system. We now employ these intuitive observations to obtain the scaling of the wave attractor. Our numerical simulations reveal that the radius of the wave attractor increases with $D$ according to a power law with an exponent of approximately $0.9$, meaning that the corresponding values of the Lyapunov function increases with this exponent. If the system size is not a multiple of $\lambda\sim D^{1/2}$ one of the three domains will be of different wavelength. Assuming the concentration of empty sites is independent of $D$ we set without loss of generality $\overline{a}=\overline{b}\sim D^{-1/2}$ and $\overline{c}\sim1-2 D^{-1/2}$. Upon inserting these concentrations into the Lyapunov function, we obtain $\mathcal{L}\sim D - 2 D^{3/2}$, which is to leading order in agreement with our results~\footnote{The reason for the slight deviation of the numerical data from the linear scaling might be that the wave attractor is slightly asymmetric. The value of the Lyapunov function is therefore not an optimal indicator for its radius.}.

\begin{figure}[tb]
\begin{center}
\includegraphics[width=0.32\columnwidth]{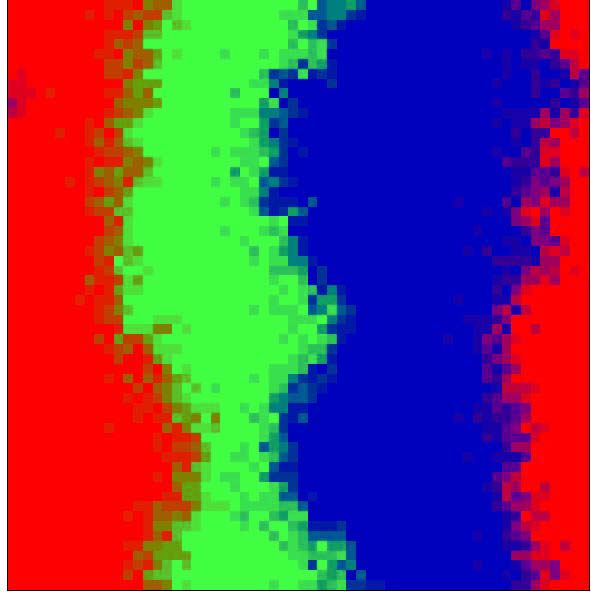}
\includegraphics[width=0.32\columnwidth]{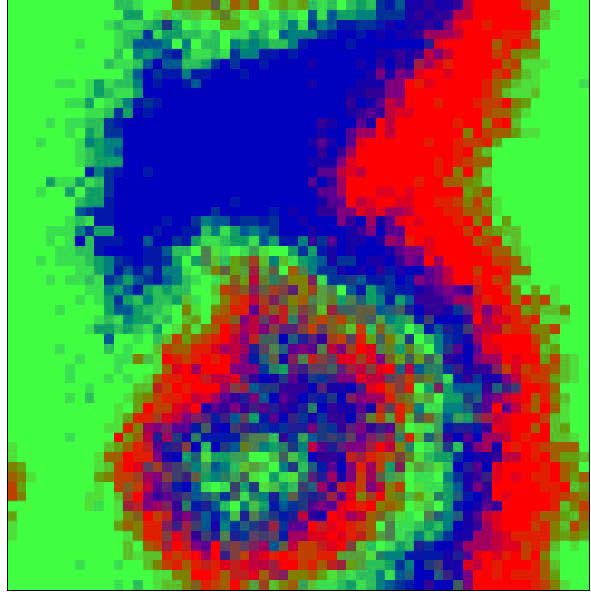}
\includegraphics[width=0.32\columnwidth]{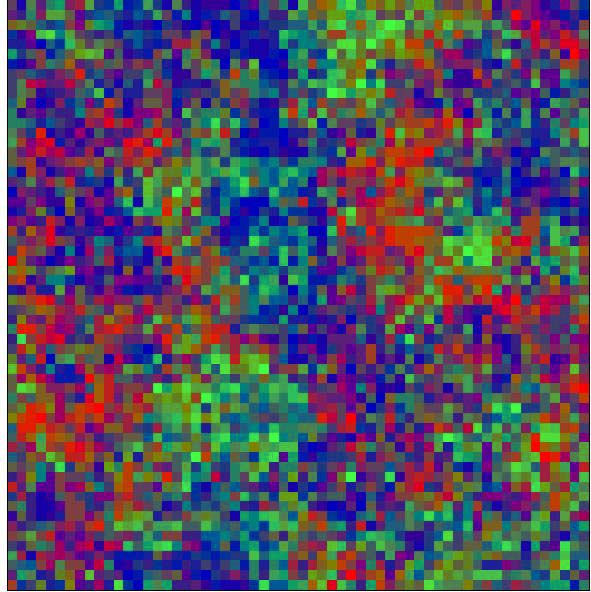}
\caption{(Color online) Illustration of the instability of wave fronts in the cyclic Lotka-Volterra system. The initial condition at $t=0$ was chosen as three domains of equal size in cyclic order. The pictures show snapshots at times $t=135$, 180, and 300. Color (gray scale) denotes species concentrations as described in Fig.~\ref{fig:snapshots_ml}. Parameters were $D = 10^{-4}$, $M = 8$ and $L = 80$.\label{fig:waves_lv}}
\end{center}
\end{figure}

\changed{Summarizing, we find that the spatio-temporal dynamics changes qualitatively at certain threshold values of the diffusion constant.  These changes are finite size effects in the sense that they arise as a result of the comparison of certain length scales. We use the term ``transition'' for this behavior in the sense that macroscopic properties of the system change qualitatively and abruptly at these threshold values. This is particularly evident in the mean first passage times to extinction. In the language of nonlinear dynamics the system undergoes bifurcations as a function of the mobility. }

\section{The cyclic Lotka-Volterra Limit \label{sec:lv}}
In the limit $\sigma\rightarrow 0$, $\mu\rightarrow 0$ only reactions remain, where the replication of predators does not require the availability of empty spaces. The resulting model is then of the Lotka-Volterra type \cite{Lotka1920,volterra-1926-31}, and characterized by a reduced set of chemical reactions:
\begin{eqnarray}
AB\stackrel{\nu}{\rightarrow}AA,\; & BC\stackrel{\nu}{\rightarrow}BB,\; & CA\stackrel
{\nu}{\rightarrow}CC\, .
\label{eq:rr_lv}
\end{eqnarray}
This model is often referred to as the three-species Lotka-Volterra model. Although at a first glance there are no dramatic differences to the May-Leonard reactions, Eq.~(\ref{eq:rr_ml}), the ensuing nonlinear dynamics is vastly different. The deterministic rate equations read
\begin{eqnarray}
\partial_t a &=&\nu a (b-c) \nonumber\, ,\\
\partial_t b &=&\nu b (c-a) \label{eq:re_lv}\, ,\\
\partial_t c &=&\nu c (a-b) \nonumber\, .
\end{eqnarray}
Without loss of generality, we also fix the normalization of total concentrations: $a+b+c=1$. The nonlinear dynamics of the well-mixed cyclic Lotka-Volterra model again exhibits the same absorbing fixed points as the general model (\ref{eq:rr}). The reactive fixed point is now given by 
\begin{equation}
(a^*,b^*,c^*)= \biggl( \frac13,\frac13,\frac13 \biggr) \, .
\end{equation} 
It is, however, neutrally stable as the real parts of the eigenvalues, Eq.~(\ref{eq:eigenvalues}), are zero. In fact, $\dot{\mathcal{L}}=0$ for any $a$, $b$ and $c$, such that starting from any point on the phase plane, the trajectories form neutrally stable cycles. 

Similar to the May-Leonard model, species' mobility drastically alters the system's collective dynamics~\cite{Reichenbach2008b}. However, the ensuing spatio-temporal dynamics of the cyclic Lotka-Volterra and the May-Leonard model differ qualitatively. This behavior can be understood upon considering the dynamics of domain boundaries separating different species. In the May-Leonard model the separation of selection and reproduction processes is counteracting the roughening of these domain boundaries due to stochastic fluctuations: If a species from one domain enters the other species' domain, it first creates empty sites. Since these empty sites are occupied with a higher probability by offsprings of individuals from the invaded species rather than by invaders, the invasion process is unlikely to be successful. This stabilizes spatially separated domains in the May-Leonard model. In contrast, in the cyclic Lotka-Volterra model an invader directly replaces the invaded species such that it has a higher probability of success. As illustrated in Fig.~\ref{fig:waves_lv}, this leads to a roughening instability of planar wave fronts.  However, this does not imply the total loss of any spatial correlations. To the contrary, there are still strong correlations and they play a fundamental, yet subtle, role in the spatio-temporal dynamics and the processes leading to the extinction of all but one species.

In our simulations we observe different dynamic processes depending on the mobility. For large diffusion constants, where the system can be considered well-mixed, we recover the homogeneous oscillations as predicted by the rate equations~\eqref{eq:re_lv}. We still find homogeneous oscillations if the mobility is decreased. However, as we will see below, these systemwide oscillations are of entirely different nature as the neutrally stable orbits found in the well-mixed system. For even lower mobility we finally find a seemingly random appearance and dissolution of spatial clusters. These clusters are convectively unstable spiral waves, which, due to a roughening transition associated with an Eckhaus instability, appear, move and annihilate. 

\subsection{Extinction times and extinction probabilities}

\begin{figure*}[tb]
\begin{center}
\includegraphics[width=0.9\textwidth]{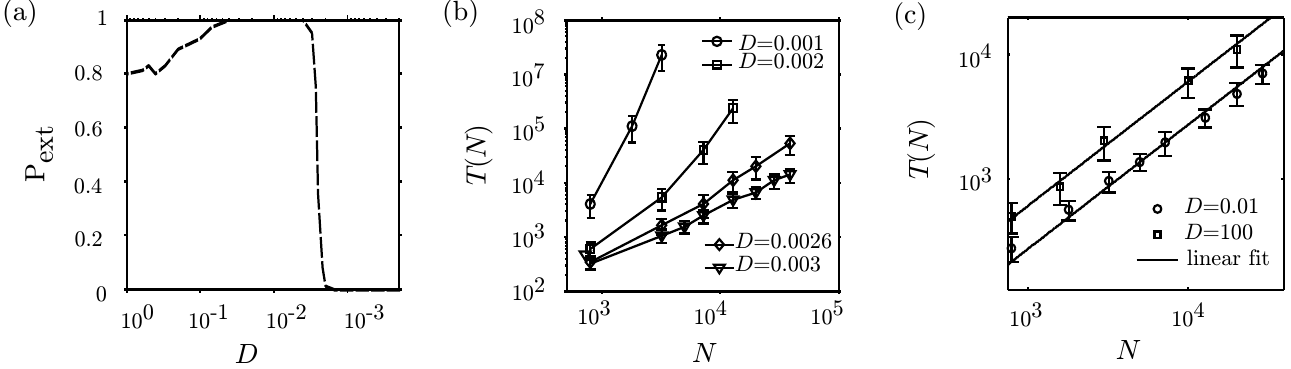}
\caption{(a) Probability that the system with Lotka-Volterra reactions reaches an absorbing state before $t=N$. We observe a sharp transition from survival to extinction at $D\approx 3\cdot 10^{-3}$. In the well mixed limit the extinction probability converges to a finite value of 0.8 ($M=8$, $L=60$).  (b) and (c) show the scaling of the mean first passage time into any of the absorbing states with the system size $N$. Two phases can be identified. For $D<3\cdot 10^{-3}$ the scaling becomes exponential, hinting at an escape process from a metastable state. In the well mixed case ($D=100$) the scaling is linear, in agreement with the escape out of a neutrally stable state. In the intermediary regime our results are in agreement with both a logarithmic and a linear scaling.
}
\label{fig:lifetimes_lv}
\end{center}
\end{figure*}

As discussed previously~\cite{Reichenbach2007a,Reichenbach2008b,Reichenbach2007}, a convenient measure to characterize the stability of the system is the probability $P_\text{ext}$ that the system has reached an absorbing state within a time proportional to the system size $N$. The simulations for our model reproduce the results found in Ref.~\cite{Reichenbach2008b}. For large $D$ our result coincides with the analytically obtained value found for the non-spatial system~\cite{Dobrinevski2012}, see Fig.~\ref{fig:lifetimes_lv}(a). For low mobilities the extinction probability is close to zero. Hence, the system is in a metastable state with extinction times scaling exponentially in the system size, cf. Fig.~\ref{fig:lifetimes_lv}(b). At a threshold value $D_c \approx 3\cdot 10^{-3}$, there is a sharp transition to $P_\text{ext} \approx 1$ indicating that extinction times scale logarithmically in the system size $N$. Indeed, Fig.~\ref{fig:lifetimes_lv}(c) indicates that the scaling of mean first passage times is sublinear. For even larger values of the diffusion constant, the extinction probability decreases again until it reaches a value of $0.8$ in the well-mixed limit. Here, extinction times scale linearly, as demonstrated by Fig.~\ref{fig:lifetimes_lv}(c). Hence, the global dynamics is characterized by the escape out of a neutrally stable state. We conclude that spatial correlations increase the system's stability for small $D$ and destabilize it above a threshold value $D_c$.

\subsection{Global attractors and free energy landscapes}

\begin{figure}[tb]
\begin{center}
\includegraphics[width=0.99\columnwidth]{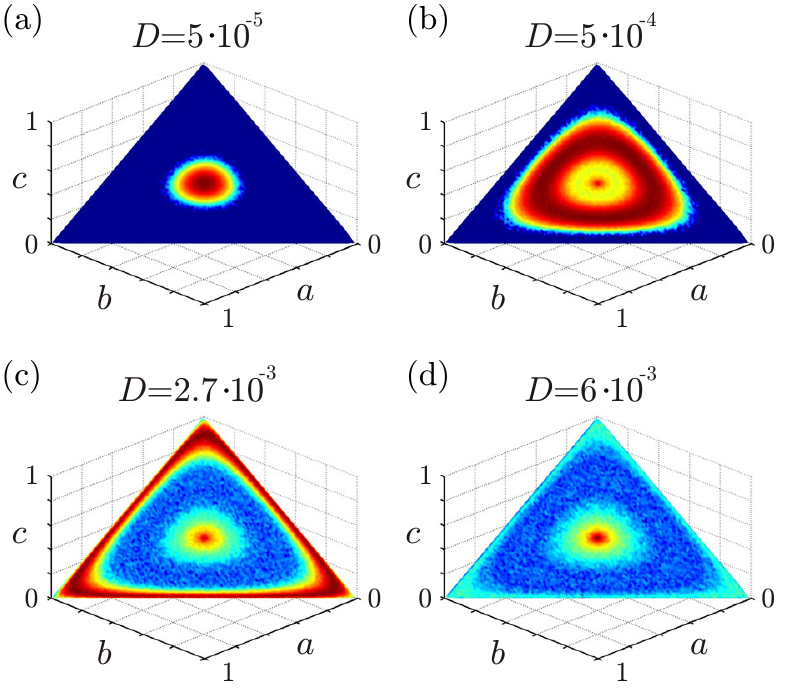}
\caption{(Color online) Probability to find the system globally in a specific state before reaching one of the absorbing states. Red (medium gray) denotes a high probability, yellow (light gray) a medium probability, and blue (dark gray) a low probability. One observes the emergence of an attracting limit cycle of the global dynamics. The attractor grows in radius with increasing $D$ and eventually reaches the boundaries of the invariant manifold. For even larger values of the diffusion constant the attractor lies outside of the invariant manifold and the global dynamics is essentially neutral. The histograms were sampled over 100 trajectories until $T=N$. Parameters were $M=8$ and $L=60$. }
\label{fig:manifolds_lv}
\end{center}
\end{figure}

As for the May-Leonard model we now employ a study of the global phase portraits to gain a deeper understanding of the ambiguous impact of spatial structures on the longevity of biodiversity. The existence of metastable states below a certain mobility threshold, suggested by the scaling of extinction times with system size, is supported by histograms of the global dynamics. Figure~\ref{fig:manifolds_lv} shows the free energy landscape projected onto the invariant manifold of Eqs. \eqref{eq:re_lv}. 
For very small values of $D$ we find an attracting region in the center of the simplex. This attractor corresponds to convectively unstable spirals. As mentioned before, smooth domain borders are subject to roughening and therefore become unstable in the Lotka-Volterra model. However, while spatial patterns can not be maintained, strong correlations exist and effectively render the global dynamics metastable. With increasing values of $D$ we observe that the trajectories describing the overall dynamics of the system are attracted towards a \emph{limit cycle}, which grows in radius and eventually reaches the boundary of the invariant manifold. This limit cycle corresponds to \emph{systemwide oscillations}. As these oscillations are linked to a metastable attractor, they are much more long-lived compared to the neutrally stable oscillations  found in the well-mixed case, \emph{i.e.} their mean lifetime scales exponentially with the system size. At some threshold value of the diffusion constant the attractor coincides with the boundary of the simplex. Then, the absorbing states are embedded within the limit cycle.  As a consequence, the global dynamics is effectively attracted towards the boundaries of the phase plane once it reaches the limit cycle's basin of attraction, and therefore rapidly reaches one of the absorbing fixed points. The global dynamics is therefore effectively heteroclinic and approaches the absorbing states exponentially fast. Hence, in this regime spatial structure destabilize the system, which explains the sub linear scaling of extinction times as shown in Fig.~\ref{fig:lifetimes_lv}(a) and (c).  For large $D$, the attractor lies outside of the simplex, such that the global dynamics on the simplex is essentially governed by neutrally stable orbits.

\begin{figure}[tb]
\begin{center}
\includegraphics[width=0.9\columnwidth]{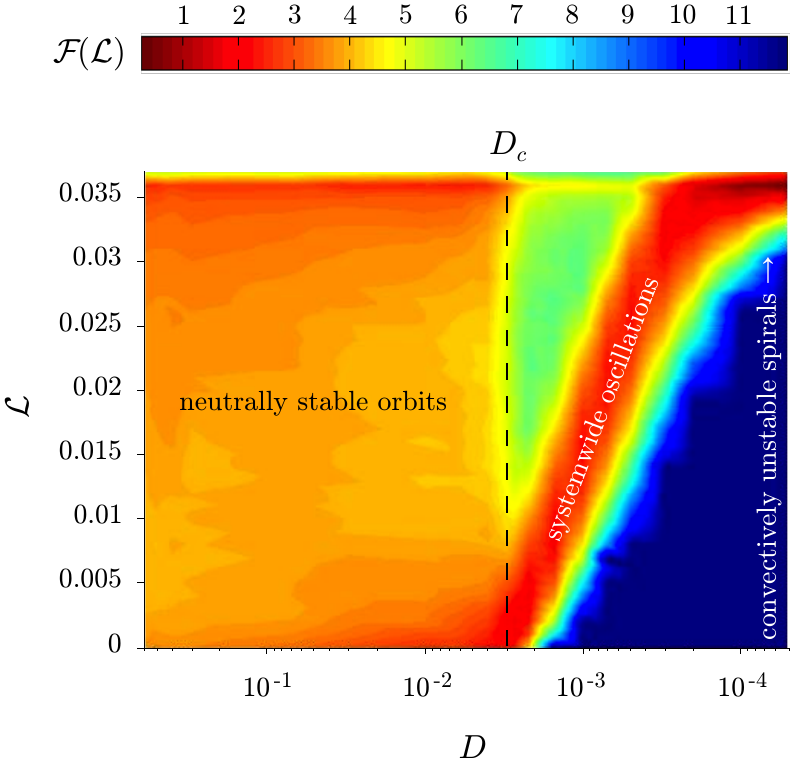}
\caption{(Color online) The global phase portrait of the Lotka-Volterra system. For each diffusion constant $D$ we plot (in colorcode or gray scale) the probability to find the system before $t = N$ at a specific value of the Lyapunov function $\mathcal{L} = \overline{a} \overline{b} \overline{c}$. Red (medium gray) denotes a high probability, yellow (light gray) a medium probability, and blue (dark gray) a low probability. Three dynamical regimes corresponding to neutrally stable orbits, systemwide oscillations and convectively unstable spirals can be identified and linked to their corresponding attractors of the global dynamics. Note that the attractor vanishes at $D \approx 3 \cdot 10^{-3}$ corresponding to the abrupt increase of extinction probabilities in Figure~\ref{fig:lifetimes_lv}.
The simulations were done with $M$ = 8 and $L = 60$, and, for numerical reasons, stopped at $T=N$. For each of roughly 20 data points in $D$ we averaged over approximately 100 trajectories. }
\label{fig:phaseportrait_lv}
\end{center}
\end{figure}

Figure \ref{fig:phaseportrait_lv} illustrates the different dynamical regimes by means of the effective free energy $\mathcal{F}$  for the cyclic Lotka-Volterra model. We identify three distinct regime, which are characterized by the shape of the effective free energy. For the well-mixed system, $D>D_c\approx3\cdot 10^{-3}$, the potential is flat and the global dynamics is neutrally stable as predicted by the rate equations~\eqref{eq:re_lv}. At $D_c$ an attractor emerges, which at this point coincides with the boundaries of the simplex ($\mathcal{L}=0$). With decreasing values of $D$ the attractor is located at increasingly large values of the Lyapunov function until at $D\approx 4\cdot 10^{-4}$ it coincides with the reactive fixed point of the rate equations~\eqref{eq:re_lv}. Comparing with our simulations we therefore identify three regimes: neutrally stable orbits, metastable systemwide oscillations, and convectively unstable spirals.  In conclusion, the behavior of the attractors of the global dynamics provides an intuitive explanation for the observed transitions in the extinction probabilities.  

\section{The intermediate regime\label{sec:int}}

While the previous sections considered important limiting cases of the reactions (\ref{eq:rr}), we now study the general case with $\sigma,\mu,\nu\neq0$. We will use the following parametrization, which allows to tune the relative weight of Lotka-Volterra type reactions and May-Leonard type reactions:
\begin{eqnarray}
\nu(s) & \equiv& s\nonumber\, , \\
\mu(s) & \equiv & 1\, , \label{eq:parametrization}\\
\sigma(s) & \equiv & 1-s\nonumber\, .
\end{eqnarray}

Here the parameter $s$ is the fraction of Lotka-Volterra type reactions, and is varied between $0$ and $1$. This choice of parametrization has two important properties: First, it conserves the limits discussed in the previous sections and makes them comparable. In the Lotka-Volterra limit and in the May-Leonard limit per time step each individual performs, on average, one active selection process or passive process, respectively. This holds for any value of $s$. Second, our simulations show that the correlation length of species concentrations stays approximately constant when changing $s$ (data not shown here). This is because in our parametrization we fix the relevant time scale, and thereby by dimensional analysis, for a given mobility, the correlation length. 

\begin{figure}[htb]
\begin{center}
\includegraphics[width=\columnwidth]{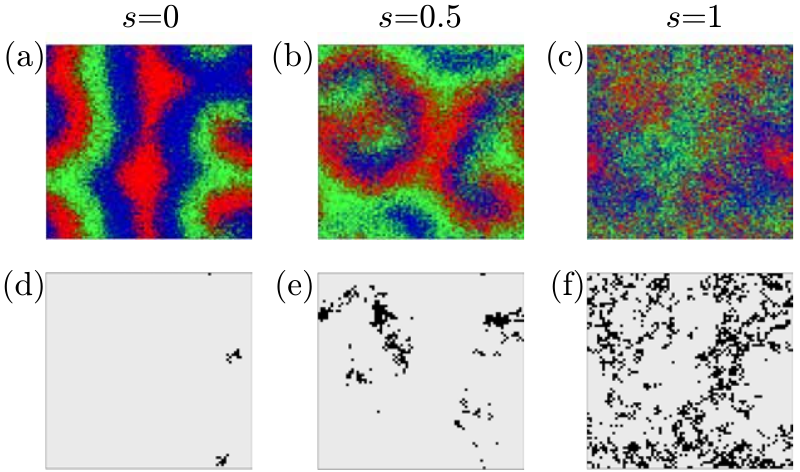}
\caption{(Color online) (a)-(c) Snapshots of the spatial distribution of species for different values of the fraction of Lotka-Volterra reactions $s$ indicated in the graph. Color denotes (gray scale) the concentration of the species $A$, $B$ and $C$, as described in Fig.~\ref{fig:snapshots_ml}. With increasing $s$ spirals become convectively unstable, \emph{i.e.} they move, annihilate and then appear again. (d)-(f) To illustrate the destabilization of spiral waves we computed for each lattice site the distance from the reactive fixed point $|\vec{y}(a,b,c)|$. Dark points show sites where $|\vec{y}(a,b,c)|$ is below a certain threshold, thereby indicating the position of spiral vertices. Parameters were $D = 10^{-4}$ (corresponding to the regime, where spirals and waves are possible in the May-Leonard model), $M = 8$ and $L = 80$.}
\label{fig:snapshots_int}
\end{center}
\end{figure}

Figure~\ref{fig:snapshots_int} shows that with increasing values of $s$ spiral patterns become convectively unstable, \emph{i.e.} the vertices start to move and annihilate. 
The destabilizing effect of Lotka-Volterra reactions on spiral patterns can also be visualized by considering the absolute value of the coordinates defined in Eq. (\ref{eq:ycoord}), $| \vec{y}(a,b,c) |$.  It gives a measure of how far the system is locally away from the reactive fixed point. Low values of  $|\vec{y} |$ correspond to a locus where each species is present at approximately equal concentrations, and therefore  indicate the position of spiral vertices. In Fig.~\ref{fig:snapshots_int}(d)-(f) black dots correspond to positions, where this absolute value is smaller than 0.13 \footnote{The threshold value was chosen to optimize the accentuation of the spiral vertices.}.  We thus infer that the spirals become unstable with increasing $s$. Indeed, the complex Ginzburg-Landau equation (\ref{eq:cgle}) predicts an Eckhaus instability, implying that the spirals vertices become convectively unstable~\cite{Reichenbach2008b}, \emph{i.e.}  they move, annihilate and appear again constantly above a certain value of $s$. To determine this value we follow the steps given in Ref.~\cite{Aranson2002}, where the stability of planar wave solutions was studied. The waves are stable, as long as the generalized Eckhaus criterion,
\begin{equation}
1-2\frac{(1+c_3^2)Q^2}{1-Q^2}>0\, ,
\end{equation}
holds, where $Q$ is the selected wave vector, 
\begin{equation}Q=\frac{2\pi}{\lambda}\sqrt{\frac{D}{c_1(s)}}=\frac{\sqrt{1+c_3(s)^2}-1}{c_3(s)} \, .
\end{equation}
Inserting $c_3(s)$ and solving for $s$ we find a critical value of $s_E\approx 0.32$. The breakdown of stable spatial structures as the result of a roughening transition is indeed confirmed by our numerical simulations. \changed{In contrast to the transitions in $D$, the Eckhaus instability is independent of the size of spatial patterns and it can therefore be considered a transition in the strict thermodynamic sense.} It has significant, yet ambiguous, implications for the stability of biodiversity, as will be discussed in the following. 

\subsection{Extinction times}

\begin{figure}[htb]
\begin{center}
\includegraphics[width=0.99\columnwidth]{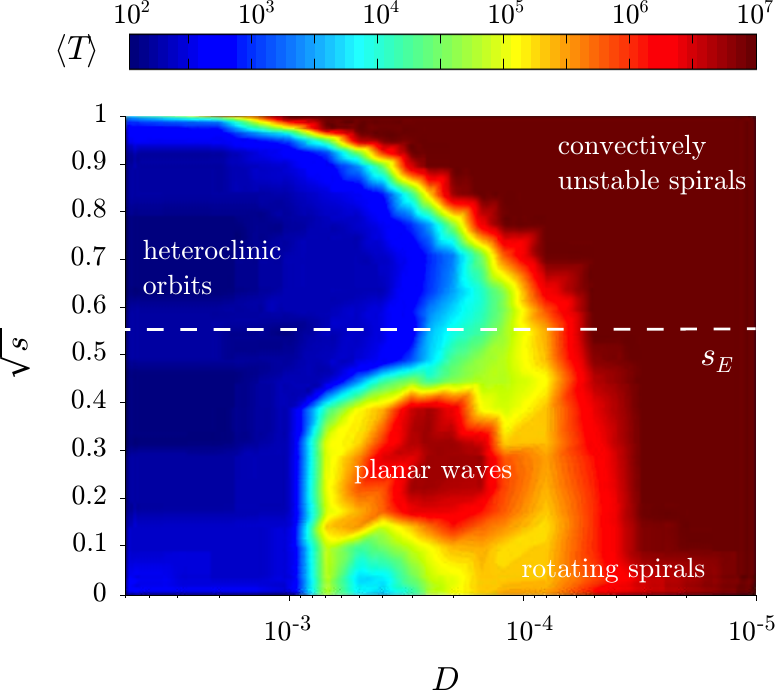}
\caption{(Color online) Average first passage time into any of the absorbing states as a function of the diffusion constant $D$ and the fraction of Lotka-Volterra reactions $s$. Red (medium gray) denotes a large lifetime, yellow (light gray) a medium lifetime, and blue (dark gray) a small lifetime. For $s=0$ we obtain the mean lifetimes shown in Fig.~\ref{fig:lifetimes_ml}. The dynamics is essentially governed by heteroclinic orbits, traveling waves and rotating spirals. With increasing $s$ the planar waves become more and more stable and dominate the dynamics for a full order of magnitude in $D$. The prominence of traveling waves leads to a local maximum in the mean lifetimes. For even higher $s$ the system undergoes an Eckhaus instability (the analytical value is denoted by a dashed line), where planar waves become unstable. The dynamics is roughly comparable to the heteroclinic orbits in the May-Leonard model. Neutral orbits are driven to the boundary of the invariant manifold by a limited fraction of May-Leonard reactions. For $s=1$ we again recover the dynamics of the Lotka-Volterra model studied in Section~\ref{sec:lv}. For each of the approximately 400 data points averages were taken over about 100 trajectories. Due to numerical constrains simulations were stopped at $T=10^7$. Parameters were $M=8$ and $L=60$.}
\label{fig:lifetimes_int}
\end{center}
\end{figure}

Figure~\ref{fig:lifetimes_int} shows the mean first passage time to one of the absorbing states as a function of $D$ and $s$. The color code as indicated in the figure is chosen such that red corresponds to large and blue to short extinction times. 
Dark red indicates the longest time simulated, $t=10^7$. The limits $s=0$ and $s=1$ correspond to the May-Leonard and Lotka-Volterra models, respectively. Varying $s$, however, does not simply interpolate between these two limits, but leads to a rather rich and complex dynamics. In particular, there is a local maximum in the mean extinction time for finite values of $s$ below $s_E$. We infer from our simulations that this maximum is linked to the emergence of planar traveling waves. In contrast to the May-Leonard model ($s=0$), planar waves seem to be increasingly important in this regime: They dominate the dynamics for a rather broad range in the diffusion coefficient. Moreover, they seem to be much more stable as compared to the May-Leonard case, which can be seen by comparing  Fig.~\ref{fig:lifetimes_ml} and Fig.~\ref{fig:lifetimes_int}. While the exact reason for this remains unclear, the stabilization of planar waves seems to be related to a change in the wavelength and thereby a reduction in the oscillations of the global concentrations. This can be inferred from the global phase portraits, as discussed below. For small $D$, we again find metastable rotating spirals. For the well-mixed system we find short first passage times. The concentrations there perform homogeneous oscillation, which we identified with heteroclinic orbits of the global trajectories for the May-Leonard case $s=0$. These orbits cover a broad parameter regime. In particular, they also arise for values of $s$, where most of the reactions are of Lotka-Volterra type. The reason for this can be inferred from the stability of the reactive fixed point of the rate equaltions~\eqref{eq:re}. The corresponding eigenvalues \eqref{eq:eigenvalues} retain a non vanishing positive real part. The trajectories of the global dynamics are therefore driven to the vicinities of the absorbing points exponentially fast. As a result, even for $s\approx 0.9$ the global dynamics is determined by a tiny fraction of May-Leonard reactions.

\begin{figure*}[p]
\begin{center}
\includegraphics[width=0.99\textwidth]{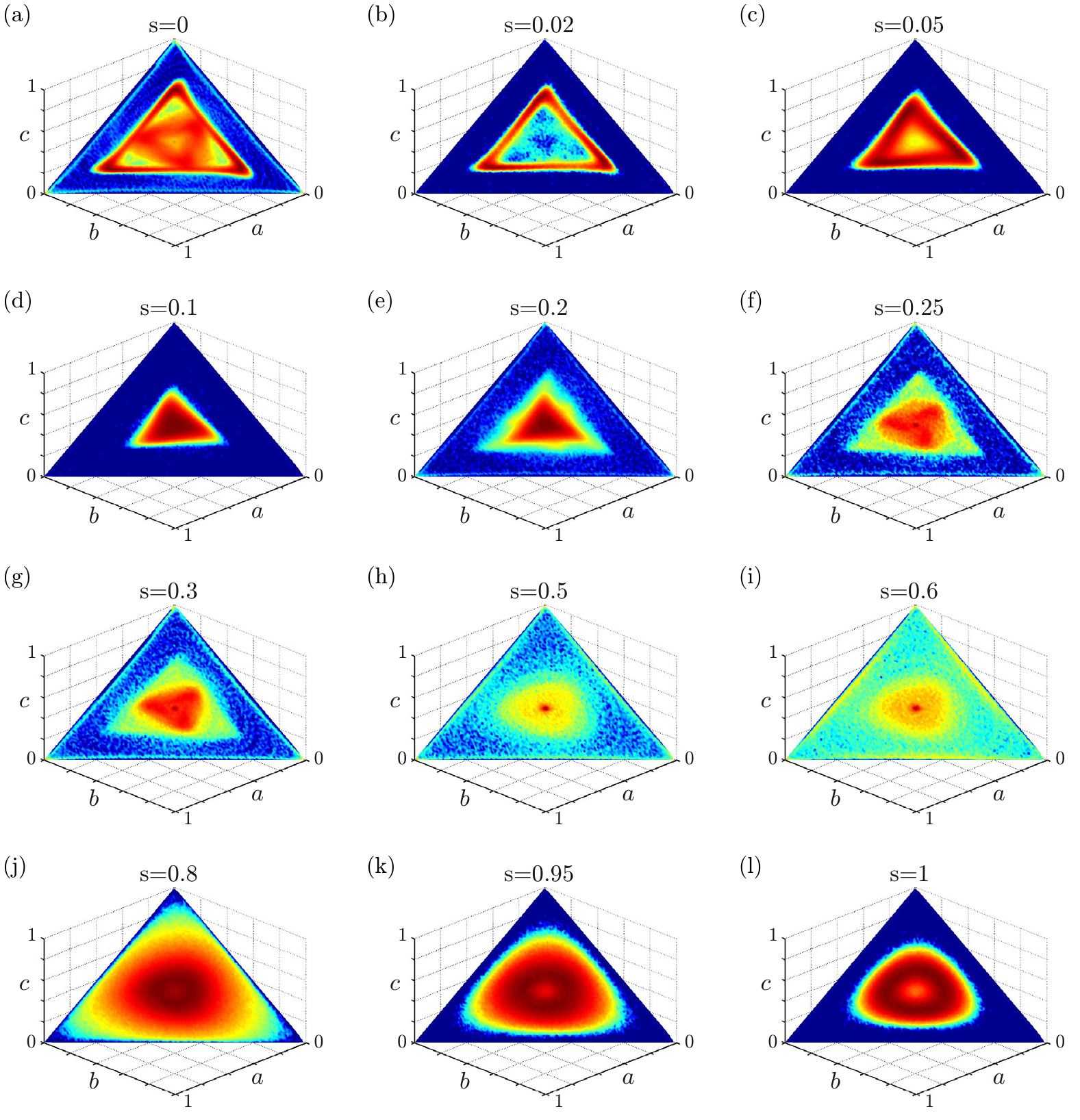}
\caption{(Color online) Probability to find the system in a specific state before reaching one of the absorbing fixed points for fixed $D=3\cdot 10^{-4}$.  The histogram is projected onto the invariant manifold of the rate equations (\ref{eq:re}). We varied the fraction of Lotka-Voltera reactions from $s=0$ (top left) to $s=1$ (bottom right). Starting at the classic May-Leonard model ($s=0$), where attractors for planar waves and rotating spirals can be identified, the attractor for the spirals disappears with growing $s$. The remaining attractor contracts to the reactive fixed point and also, when the system undergoes an Eckhaus instability, dissolves. For an even larger fraction of Lotka-Volterra reactions the global dynamics is driven outward by a limited fraction of May-Leonard reactions and is comparable to the heteroclinic orbits found in the May-Leonard model. When a majority of the reactions are of Lotka-Volterra type, \emph{i.e.} for $s$ not much smaller than 1, we again observe the emergence of an attracting limit cycle corresponding to systemwide oscillations. Parameters where $M=8$ and $L=60$.}
\label{fig:manifolds1_int}
\end{center}
\end{figure*}

\changed{The roughening transition is complicated by threshold values in $D$, corresponding to the onset of planar waves and spirals, and the dissolution of the former. These threshold values take the same values as in the limiting case of only May-Leonard reactions.}
As the value of $s$ exceeds the roughening transition (Eckhaus instability) we observe a sharp transition between long extinction times for small values of $D$ and short extinction times for large values of $D$. In the latter regime, spirals and planar waves become convectively unstable as predicted by the complex Ginzburg-Landau equation (\ref{eq:cgle}). For spiral waves this is illustrated by Fig.~\ref{fig:snapshots_int}. Nevertheless, strong correlations exist and mean times to extinction are large in this regime. From our simulations we infer that the dominant dynamic process in this regime can be identified as the convectively unstable spirals also found in the Lotka-Volterra limit. Note, however, that due to a truncation of simulation times not all details may be resolved in this regime.

\subsection{Effective free energy and Lotka-Volterra limit}

To study how the Lotka-Volterra limit is reached, we computed the effective free energy $\mathcal{F}$ as a function of $\mathcal{L}$. We focus on the case $D=3\cdot 10^{-4}$, which entails the regime of stable planar waves, cf. Fig.~\ref{fig:lifetimes_int}.
In the May-Leonard model this corresponds to the regime shortly below the lower critical point in the diffusion constant, where the wave attractor and the spiral attractor coexist. Figure~\ref{fig:manifolds1_int} demonstrates that the observed changes in extinction times are related to the emergence, disappearance and changes in the characteristics of attractors of the global dynamics. The limit of the May-Leonard model ($s=0$) was already discussed in Section \ref{sec:ml}. Attractors for rotating spirals and planar waves are visible. When the fraction of Lotka-Volterra reactions is slightly increased the spiral attractor disappears while the wave attractor remains, see Fig.~\ref{fig:manifolds1_int}(b). The latter shrinks in size, hinting at an increasing wave length [Figs.~\ref{fig:manifolds1_int}(c)-(e)]. The attractor then contracts towards the reactive fixed point[Figs.~\ref{fig:manifolds1_int}(d)-(f)]. In Fig.~\ref{fig:lifetimes_int} this regime corresponds to the local maximum in extinction times. At the point, where the system undergoes an Eckhaus instability, the attractor dissolves [Fig.~\ref{fig:manifolds1_int}(g) and (h)]. The dynamics is then dominated by global oscillations which are driven outward by a limited number of May-Leonard reactions [Fig.~\ref{fig:manifolds1_int}(h) and (i)]. This regime is therefore closely related to the heteroclinic orbits found in the May-Leonard model. For an even larger fraction of Lotka-Volterra reactions a new attracting limit cycle emerges, corresponding to the systemwide oscillations found in Section \ref{sec:lv} [Fig.~\ref{fig:manifolds1_int}(j)-(l)].

\begin{figure}[h!]
\begin{center}
\includegraphics[width=0.99\columnwidth]{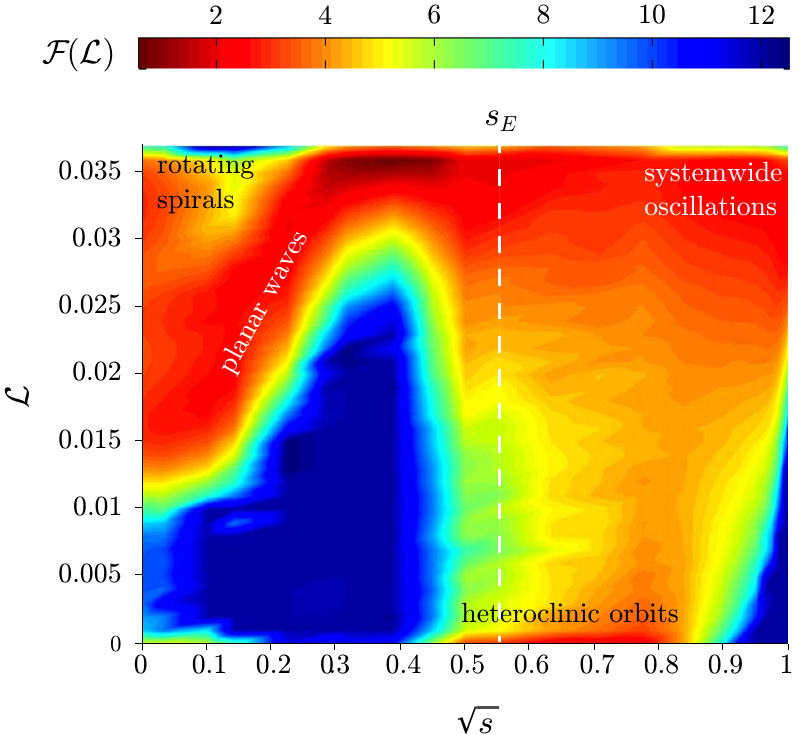}
\caption{(Color online) To study the most intriguing line of Fig.~\ref{fig:lifetimes_int}, $D=3\cdot 10^{-4}$, in more detail we computed the effective free energy $\mathcal{F}$ at specific values of the Lyapunov function $\mathcal{L}$ for different values of $s$. Red (dark gray) denotes minima of the potential and thereby attractors of the global dynamics. Yellow (light gray) signifies intermediate values and blue (dark gray) large values of the effective free energy. We identify several regimes depending on the relative strength $s$ of the different types of competition. For $s=0$ we recover the coexisting wave and spiral attractors of the May-Leonard model. With increasing values of $s$ only the wave attractor remains and approaches the reactive fixed point of the global dynamics ($\mathcal{L}=0.037$). At the Eckhaus instability (dashed line) the wave attractor dissolves. Instead, an attractor corresponding to global heteroclinic orbits emerges. Only, when almost all reactions are of Lotka-Volterra type an attractor close to the reactive fixed point emerges. The latter corresponds to the limit cycle found in the cyclic Lotka-Volterra model, see Sec.~\ref{sec:lv}. Comparing with our simulations we find that these attractors are linked to rotating spirals, planar waves, global, heteroclinic orbits and systemwide oscillations. Simulation parameters were $M=8$ and $L=60$.}
\label{fig:phaseportrait1_int}
\end{center}
\end{figure}

The results are summarized in a free energy landscape as a function of $s$, cf. Fig.~\ref{fig:phaseportrait1_int}. 
For $s=0$ we find the attractors of the planar and spiral waves of the May-Leonard model, cf. Fig.~\ref{fig:phaseportrait_ml}. When the fraction $s$ of Lotka-Volterra reactions is increased the attractor of planar waves shrinks to the center of the manifold. As a result, there are no oscillations in the overall densities, which is in contrast tothe May-Leonard model, where these oscillations stem from waves having a wavelength close but unequal to the system size. As a result of the lack of oscillations, planar waves become increasingly stable in this regime. At the Eckhaus instability, $s_E$, spatial patterns become unstable. The dynamics can then be best described as heteroclinic orbits. The system globally performs orbits, which are driven to the boundary of the manifold by the reactions of May-Leonard type. Hence, even a tiny fraction of May-Leonard reactions determines the global dynamics in this regime. \changed{This is not surprising, as the conservation law associated with the cyclic Lotka-Volterra model holds precisely only in the case $s=1$.} For values of $s$ close to 1 reactions involving empty sites become unimportant. We then find the attractor corresponding to systemwide oscillations, cf. Fig.~\ref{fig:phaseportrait_lv}. Summarizing, in the model with direct and indirect competition we find a surprisingly rich variety of dynamic processes affecting the longevity of biodiversity in a much more complex manner than one would naively expect from an Eckhaus instability. In particular, we observed a local maximum in mean lifetimes if direct competition is weak but non vanishing.

\section{Conclusion}

We studied the population dynamics of three-species models where species interact with each other cyclically through both direct and indirect competition. In the limiting cases of only direct or indirect competition our model reduces to the cyclic Lotka-Volterra or May-Leonard model, respectively. For a well-mixed system, the nonlinear dynamics of these models differs significantly. While in both cases the trajectories lie on two-dimensional invariant manifolds comprising the absorbing states of extinction, their phase portraits differ qualitatively. The dynamics of the cyclic Lotka-Volterra model is characterized by neutrally stable orbits. In contrast, the dynamics of the May-Leonard model features an unstable fixed point in the center of the invariant manifold and heteroclinic orbits which approach the boundaries of the invariant manifold and hence the absorbing states exponentially fast. In the spatial versions of these models, these attractors of the well-mixed system still act locally on each lattice site. However, if one views the spatially extended system as a set of interconnected local patches, the coupling between these patches due to diffusion may lead to qualitative changes in the type and stability of these attractors.

Indeed, numerical simulations show that in spatially extended systems there is a rich diversity of spatio-temporal patterns depending on the systems' parameters. The goal of this work was to identify and characterize the dynamic processes responsible for the transient maintenance of biodiversity and ultimately leading to extinction in the spatial models. To this end, we investigated the phase portrait of the overall concentrations for the species comprising the system and analyzed the ensuing global attractors of the dynamics and how they are connected with the different types of spatio-temporal dynamics. Moreover, based on a statistical analysis of the system trajectories on the global phase portrait, we defined an effective free energy which gave us valuable information about the scaling of extinction times with the system size. In particular, the minima of the free energy correspond to metastable dynamical processes.

In the limit corresponding to the spatial May-Leonard model, the minima in the effective free energy landscape of the global phase portrait are linked to three distinct spatio-temporal patterns: (i) spatially homogeneous oscillatory behavior, (ii) planar traveling waves and (iii) rotating spirals. Importantly, the characteristics of the global attractors change qualitatively at certain threshold values of the mobility. This means that the length scales associated with the spatial patterns changes, which affects their stability and thereby the probability to find the system in such a state. In particular, below an upper threshold value of the mobility a triangular attractor corresponding to traveling waves emerges. This attractor can be regarded as a limit cycle of the global dynamics. It grows in size for decreasing mobility, reflecting a decreasing wavelength. At a lower threshold value of the mobility  a second limit cycle of the global concentrations is found, which is located inside the attractor of the traveling waves.  There, rotating spirals emerge. In this regime we observe the coexistence of two dynamic processes, planar waves and rotating spirals. Which of the two dynamic regimes is actually realized is determined by stochasticity and the initial conditions.  For even lower mobility, the attractor of the traveling waves dissolves, as the correlation length is too small compared to the system size to maintain planar domain borders. In this regime only the attractor of rotating spirals remains, which dominates the dynamics predominantly.  

As opposed to this behavior, in the limit of reactions of Lotka-Volterra type only, the system does not exhibit stable spatial patterns. However, there are strong spatial correlations, and they manifest themselves in an attractor of the global dynamics. This attractor has the form of a rounded triangle around the reactive fixed point and corresponds to a limit cycle of the global concentrations. The radius of the limit cycle grows with increasing mobility: when the attractor, with increasing mobility, reaches the boundaries of the invariant manifold the dynamics passes from metastability (exponential scaling of extinction times with the system size) to rapid extinction (sub-linear scaling of extinction times with the system size). For even larger mobilities, the radius of the global limit cycle outgrows the boundaries of the simplex. The mean time to extinction then scales linearly with the system size. Hence, in contrast to the May-Leonard model a single attractor here is responsible for three distinct scaling regimes. 

Finally, we found a remarkably complex behavior when varying the relative strengths of direct (Lotka-Volterra) and indirect (May-Leonard) competition. If direct competition is weak compared to indirect competition, planar traveling waves are an increasingly important constituent of the extinction dynamics as compared to the May-Leonard case. These planar waves are very stable, leading to a local maximum of extinction times in the phase diagram. Simultaneously, we observe that in contrast to the May-Leonard model rotating spirals do not form for intermediary mobilities. This is reflected in the dissolution of the corresponding attractor. At a specific fraction of Lotka-Volterra reactions the system undergoes an Eckhaus instability: traveling waves and rotating spirals become unstable. The Eckhaus instability manifests itself in the vanishing of the attractors of planar waves and rotating spirals. Beyond the Eckhaus instability, a new attractor emerges, corresponding to global oscillatory behavior for high mobility and convectively unstable spirals for low mobilities. Summarizing, we find that the spatia-temporal dynamics of cyclic populations models with both, direct and indirect competition is surprisingly rich and differs qualitatively from the cyclic Lotka-Volterra and May-Leonard models. We identified several threshold values of the mobility and the relative strength of the two types of competition. 

In conclusion, scaling of extinction times with the system size change abruptly at certain threshold values of the mobility and the relative strength of the two types of competition. We showed that the dynamic processes leading to the transient maintenance of biodiversity are linked to attractors of an effective free energy of the overall concentrations. The characteristics of these attractors change upon certain threshold values, thereby giving insight into the mechanisms underlying these phase transitions. By means of extensive numerical simulations we provide the complete phase diagrams, which are rationalized by scaling arguments based on properties of the complex Ginzburg-Landau equation.

\changed{We believe that the method of global phase portraits and the ensuing effective free energy landscapes (renormalized reaction terms) might also give a deeper insight into the dynamics of spatial ecological models and reaction-diffusion systems in other fields of biology. In particular, further studies may apply this method to understand  epidemic models, asymmetric four species models or more complex food webs.}

\begin{acknowledgments}
This research was supported by the German Excellence Initiative via the program  `Nanosystems Initiative Munich' and the German Research Foundation via contract FR 850/9-1. We thank David Jahn, Johannes Knebel, Markus Weber, and Tobias Reichenbach for fruitful discussions.
\end{acknowledgments}

%\bibliographystyle{prsty}

%\bibliography{./PRE_2012}

\end{document}